\documentclass{aa}

\usepackage{lscape}
\usepackage{natbib}
\usepackage{graphicx}
\usepackage{hyperref}

\begin{document}

\title{The X-ray/SZ view of the virial region\\
\Large{I. Thermodynamic properties}}

\author{D. Eckert\inst{1} \and S. Molendi\inst{2} \and F. Vazza\inst{3,5}  \and S. Ettori\inst{4,5} \and S. Paltani\inst{1}}
\institute{
Astronomical Observatory of the University of Geneva, ch. d'Ecogia 16, 1290 Versoix, Switzerland\\
\email{Dominique.Eckert@unige.ch}
\and
INAF - IASF-Milano, Via E. Bassini 15, 20133 Milano, Italy
\and
Hamburg Observatory, Gojansbergsweg  112, 21029 Hamburg, Germany
\and
INAF - Osservatorio Astronomico di Bologna, Via Ranzani 1, 40127 Bologna, Italy
\and
INFN, Sezione di Bologna, viale Berti Pichat 6/2, 40127 Bologna, Italy
}

\abstract{}{We measure the thermodynamic properties of cluster outer regions to provide constraints on the processes that rule the formation of large scale structures.}{We derived the thermodynamic properties of the intracluster gas (temperature, entropy)  by combining the Sunyaev-Zel'dovich thermal pressure from \emph{Planck} and the X-ray gas density from \emph{ROSAT}. This method allowed us to reconstruct for the first time temperature and entropy profiles out to the virial radius and beyond in a large sample of objects.}{At variance with several recent \emph{Suzaku} studies, we find that the entropy rises steadily with radius, albeit at at a somewhat lower rate than predicted by self-similar expectations. We note significant differences between relaxed, cool-core systems and unrelaxed clusters in the outer regions. Relaxed systems appear to follow the self-similar expectations more closely than perturbed objects.}{Our results indicate that the well-known entropy excess observed in cluster cores extends well beyond the central regions. When correcting for the gas depletion, the observed entropy profiles agree with the prediction from gravitational collapse only, especially for cool-core clusters.}
\titlerunning{The X-ray/SZ view of the virial region I}
\maketitle

\section{Introduction}

The outskirts of galaxy clusters are characterized by ongoing accretion processes and significant departures from virialization and hydrostatic equilibrium. Still, at the present epoch they host the transition between regular large-scale accretions and the injection of shocks and turbulent motions, and they are the sites where the transition between large-scale infall motions and the dense and hot ICM occurs. Departures from hydrostatic equilibrium in the form of turbulence motion \citep[e.g.,][]{dolag05,vazza11a} and non-thermal sources of energy such as cosmic rays \citep[e.g.,][]{vazza12a,pfrommer07} and magnetic fields \citep{dolag99,bruggen05} are expected to play a more important role in these regions than in cluster cores, and the infall of smaller structures along large-scale filaments may cause the material in these regions to be clumpy \citep{nagai,simionescu} and asymmetric \citep{vazzascat,e12}. These effects may bias the measurements of cluster masses using X-ray and SZ proxies \citep{rasia,nagai07,piffaretti,burns10}, thus limiting the use of galaxy clusters as high-precision cosmological probes \citep{allen11}. 

For all these reasons, a deeper understanding of the state of the intracluster gas in cluster outskirts is required. However, because of the very low surface brightness of the X-ray signal, providing reliable temperature measurements in these regions is challenging \citep{ettoriwfxt,eckertpks}. Recently, the low background of the \emph{Suzaku} satellite allowed for the measurement of temperature profiles out to $R_{200}$\footnote{For a given overdensity $\Delta$, $R_\Delta$ is the radius for which $M_\Delta/(4/3\pi R_\Delta^3)=\Delta\rho_c$} in a handful of systems \citep[e.g.,][]{reip09,bautz,kawa,hoshino,simionescu,akamatsu,humphrey}. However, these measurements are often probing a small fraction of the total extent of these objects around the virial radius, and they may be subject to systematic problems \citep{eckertpks}. Hence, the general behavior of the temperature profiles beyond $R_{500}$ is still largely unknown. Conversely, while direct temperature measurements require high-quality X-ray spectra and an excellent knowledge of the systematic uncertainties, estimates of the (projected) gas density can be easily obtained from the flux in the soft band (0.5-2 keV), provided that the temperature does not fall below $\sim1.5$ keV. In this respect, the large field-of-view of \emph{ROSAT}/PSPC has allowed for the reconstruction of gas density profiles out to the virial radius in relatively large samples \citep{e12,vikhlinin99,neumann05}, providing a consistent indication of a steepening of the gas density profiles beyond $R_{500}$.

While the X-ray signal is proportional to the square of the gas density, and thus decreases sharply in cluster outskirts, the SZ effect \citep{sz} depends on the integrated pressure along the line of sight, which decreases more gently with radius \citep{arnaud10}. New-generation SZ instruments are now routinely detecting galaxy clusters, sometimes out to large radii \citep[e.g.,][]{plagge10,morandi,ami,sayers}. Recently, measurements of the SZ effect in a sample of 62 galaxy clusters with the \emph{Planck} satellite were reported \citep[hereafter P12]{planck5}, detecting a stacked signal as far as $\sim3R_{500}$ from the cluster center. \emph{Planck} therefore provides high-quality pressure profiles in cluster outskirts, which can be combined with X-ray information to get some important insight on the state of the intracluster medium (ICM) beyond $R_{500}$.

In this paper, we demonstrate that by combining the average SZ pressure profiles from P12 with the average \emph{ROSAT} gas density profiles presented in \citet[hereafter E12]{e12}, it is possible to accurately recover the temperature and entropy profiles in galaxy clusters, bypassing the use of X-ray spectroscopic information \citep[see also][]{ameglio07,ameglio09}. This allows us to probe the importance of additional effects (non-thermal pressure support, gas clumping, asymmetry) on the quantities recovered by assuming that the gas is in hydrostatic equilibrium. In a companion paper (hereafter Paper II), we use this information to infer the average hydrostatic mass and gas fraction profiles. Similar results are also presented for a subset of 18 individual clusters that are in common between the samples of P12 and E12.

Throughout the paper, we assume a $\Lambda$CDM cosmology with $\Omega_m=0.3$, $\Omega_b=0.05$, and $H_0=70$ km s$^{-1}$ Mpc$^{-1}$.

\section{Method}

\subsection{Basic formalism}
\label{formalism}

For this work, the two main observables are the electron pressure $P_e(r)$ from SZ measurements and the number density of electrons $n_e(r)$ from X-ray observations. The gas temperature can be obtained from these two quantities as

\begin{equation} kT(r)=\frac{P_e(r)}{n_e(r)},\label{trec}\end{equation}

\noindent with $n_e\sim1.21n_H$. Similarly, the entropy $K$ can be written as

\begin{equation} K(r)=kT(r) n_e(r)^{-2/3} = P_e(r) n_e(r)^{-5/3}. \label{entropy}\end{equation}

\noindent From the self-similar model \citep[e.g.,][]{bryan,arnaud02}, these quantities are expected to scale like
\begin{eqnarray}\noindent T & \sim & M^{2/3}h(z)^{2/3} \label{tssc}\\
K & \sim & M^{2/3} h(z)^{-2/3} \label{Kssc},\end{eqnarray}

\noindent where $h(z)=\sqrt{\Omega_m(1+z)^3+\Omega_\Lambda}$, until the accretion radius ($\sim1.5-2$ times the cluster's virial radius), and thus we rescale the profiles by these factors, such that in a purely self-similar case all profiles should match.

The radial dependence of the pressure is assumed to follow a generalized Navarro-Frenk-White (GNFW) profile \citep{gnfw,nagai07b},

\begin{equation} \frac{P(r)}{P_{500}}=\frac{P_0}{(c_{500}x)^\gamma(1+(c_{500}x)^\alpha)^{(\beta-\gamma)/\alpha}}, \label{eqgnfw}\end{equation}

\noindent where $x=r/R_{500}$, and $P_{500}$ is the expected self-similar pressure at $R_{500}$ \citep[see Appendix A of][for details]{arnaud10},

\begin{equation} P_{500}=1.65\times10^{-3} h(z)^{8/3} \left(\frac{M_{500}}{3\times10^{14}M_{\odot}}\right)^{2/3} h_{70}^2 \mbox{ keV cm}^{-3}. \label{p500}\end{equation}

\noindent In addition, the pressure profiles are rescaled by the quantity

\begin{equation} f(M)= \left(\frac{M_{500}}{3\times10^{14}M_{\odot}}\right)^{0.12}, \label{fofm}\end{equation}

\noindent which accounts for the dependence of the gas fraction on mass \citep[e.g.,][]{arnaudlxt,ettori99,vikhlinin06}. In Fig. \ref{planckprofs} and \ref{planckprofs2} of the online version we show the best-fit GNFW profiles compared to the \emph{Planck} data. The best-fit parameters are found to be always consistent with the results given in P12.

\begin{figure*}
\resizebox{\hsize}{!}{\hbox{
\includegraphics[height=5cm,width=7cm]{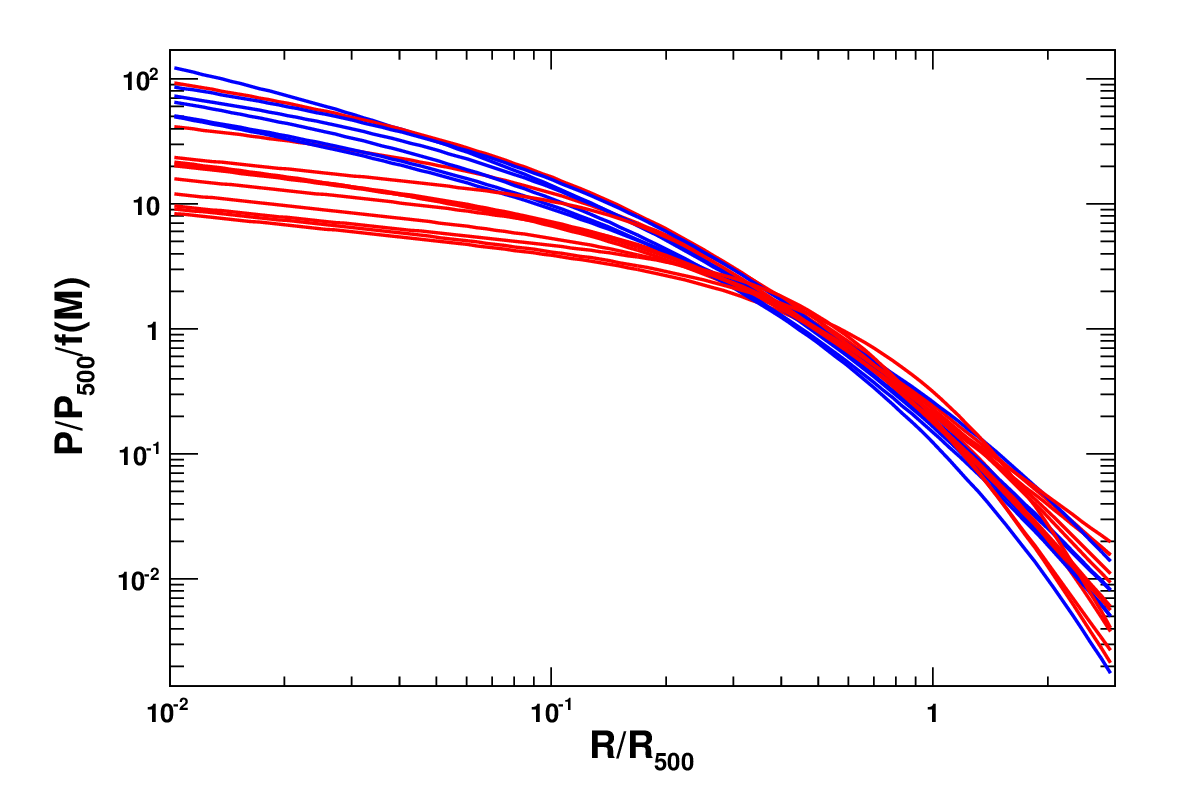}
\includegraphics[height=5cm,width=7cm]{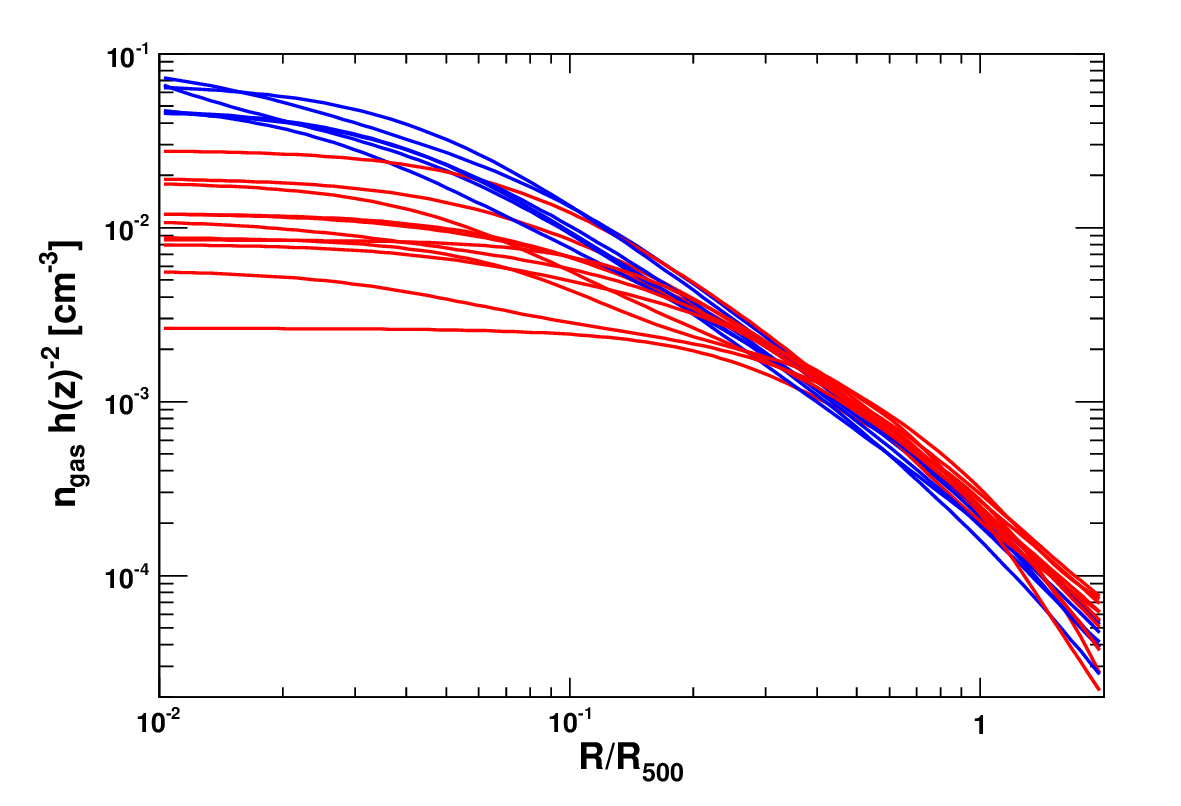}
}}
\caption{Thermal pressure (left, using the best-fit parameters from P12) and gas number density profiles (right, from E12) for the 18 objects in common between the two samples. The profiles are drawn using the functional forms defined in Eqs. \ref{eqgnfw} and \ref{vikhform}, respectively. In both cases, blue lines indicate CC systems, while NCC profiles are shown in red.}
\label{basic}
\end{figure*}

To describe the gas density profiles, we use a simplified version of the functional form given by \citet{vikhlinin06},

\begin{eqnarray} n_{gas}(r) & = & n_0\left(\frac{1}{(1+r^2/r_c^2)^{3\beta/2}}\frac{1}{(1+r^\gamma/r_s^\gamma)^{\epsilon/\gamma}}\right. \nonumber \\
\, & + &\left. \frac{R}{(1+r^2/r_{c2}^2)^{3\beta/2}} \right) \label{vikhform}, \end{eqnarray}

\noindent where $n_{gas}=n_{e}+n_{H}$. This function is projected along the line-of-sight $\ell$ assuming spherical symmetry, and is fit to the observed \emph{ROSAT} emission-measure (EM) profiles,

\begin{equation} EM(r)=1.21 \int n_H^2(\omega) \, d\ell, \mbox{ with }\omega^2=r^2+\ell^2 . \end{equation}

\noindent The parameters in Eq. \ref{vikhform} are strongly degenerate, therefore following \citet{vikhlinin06}, we fix the parameter $\gamma$ to the value of 3. We stress that this functional form can allow for both a steepening and a flattening of the density profiles, through the sign of the $\epsilon$ parameter. This function is found to adequately fit the observed emission-measure profiles in all cases ($\chi_{red}^2\leq1$). For comparison, we also used the density profiles extracted through direct deprojection as in E12 \citep{kriss}. The \citet{kriss} geometrical deprojection technique assumes that the outermost density bin is unaffected by projection effects, which leads to flatter profiles at the edge. To alleviate this problem, we applied a correction for the edge effect as explained by \citet{mclaughlin99,buote00}. After applying this correction, the parametric and deprojected density profiles are found to match very well, showing differences that usually do not exceed 10\% at all radii. The comparison between the deprojected and parametric density profiles can be found for each individual cluster in Fig. \ref{rosatprofs} and \ref{rosatprofs2} of the online material.

In Fig. \ref{basic} we show the pressure (left) and density profiles (right) for a subset of 18 clusters which are in common between P12 and E12. Based on their central entropy \citep{cavagnolo}, six of these clusters are classified as cool-core (CC, $K_0<30$ keV cm$^2$), while the remaining 12 are non-cool core (NCC). The list of these clusters is given in Table \ref{tab1}. We note that in our sample, a detection of the emission beyond $R_{200}$ at a significance of at least $2\sigma$ is achieved for only three CC clusters, so the average CC profiles may not be representative in this radial range. To rescale the density profiles, we used the values of $R_{500}$ given in \citet{planck11}. For the general properties of these clusters, we refer to \citet{planck11} and E12. 

\subsection{Monte-Carlo Markov chain estimation of errors and confidence intervals}
\label{sec:mcmc}

Providing accurate and meaningful error estimates and confidence intervals is crucial to probe the state of the ICM in cluster outskirts, where both the X-ray and SZ signal are weak. We adopt here a Monte-Carlo Markov chain (MCMC) technique to compute the confidence envelope of the functional forms describing the gas density (Eq. \ref{vikhform}) and the pressure (Eq. \ref{eqgnfw}). MCMC \citep[e.g.,][]{gilks} is the method of choice to probe the error distribution of parameters in fitting multi-dimensional functions. It is particularly suitable to cases where the parameter space is complex and parameters are strongly correlated. MCMC also allows one to make any statistical inference on the basis of the distributions it has sampled. 

We developed an implementation of MCMC based on the Metropolis-Hastings algorithm (formally Metropolis algorithm, because the Hastings ratios were always 1). Chains were calculated for all parameters of the pressure and density profiles for all 18 objects and for the average profiles. Gaussian proposals were used, taking into account the one-dimensional errors on each parameter, which were rescaled to reach a rejection ratio between 0.5 and 0.8 for each chain. Each of the 38 chains contains $10^6$ steps. No burn-in phase was used, because the chains started from the best fit; however, we checked that discarding the first $10^5$ steps had no effect on the chain distributions. Since we are mostly interested in the cluster external regions, the procedure was applied only on the data beyond $0.2R_{500}$. The parameters affecting only the central regions ($\le 0.1 R_{500}$) were fixed to their best-fit values during the MCMC procedure; these parameters are $\gamma$ for the pressure profiles, which was fixed to the value of 0.31 as in P12, and $r_{c2}$ and $\beta$ for the density profile. Since for both functional forms the parameters are strongly correlated, this allows one to explore the parameters relevant to the outer regions in more detail. In the case of the average NCC pressure profile, fixing the inner slope $\gamma$ to the value of 0.31 leads to an underestimation of the data points by 10-15\% at all radii beyond $0.2R_{500}$. Since our focus for this work is on the outer regions, in this specific case we allowed this parameter to vary during the chain, which gives a better description of the data beyond $0.2R_{500}$. We checked the convergence of the chains by running a few of them for $2\times10^6$ steps and comparing the distributions of the short and long chains.

\begin{figure}
\resizebox{\hsize}{!}{\hbox{
\includegraphics[height=5cm,width=7cm]{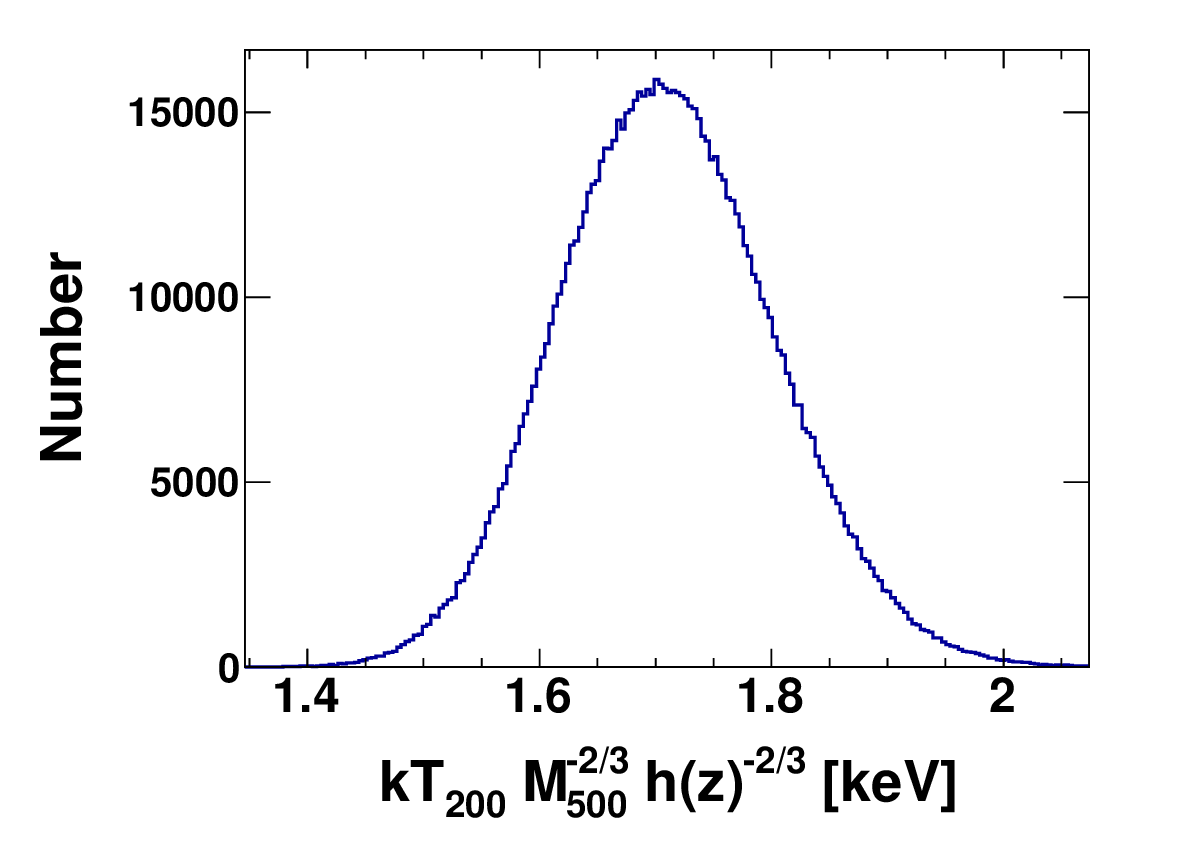}
\includegraphics[height=5cm,width=7cm]{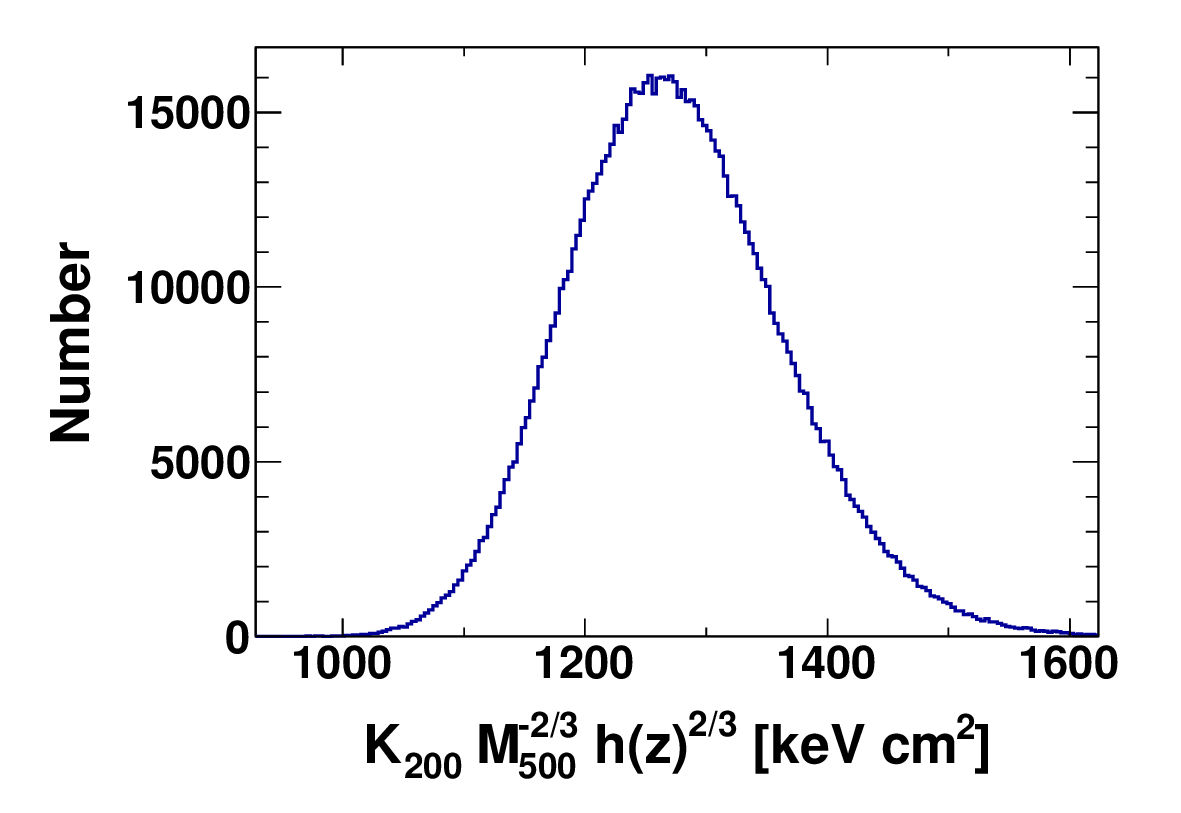}
}}
\caption{Posterior distributions for the reconstructed temperature (left) and entropy (right) at $R_{200}$ from the average pressure (from P12) and density profile (from E12) after $10^6$ MCMC steps.}
\label{mcmc}
\end{figure}

Uncertainties in the fit parameters can be straightforwardly propagated in the determination of the temperature and entropy profiles using the chains, allowing one to determine the confidence intervals on these profiles. For the average profiles and for each object, the temperature and entropy profiles were constructed for each step of the corresponding chains. The confidence interval at a given radius is then given by the distribution of the profile values at this radius. For example, Fig. \ref{mcmc} shows the error distributions for the temperature (left) and entropy (right) at $R_{200}$ from the average pressure and EM profiles built using MCMC. 

\section{Results}

\subsection{Validation of the method}

To validate our method for the reconstruction of gas temperatures, we used temperature measurements from the literature \citep{lm08,cavagnolo,xmmcat} and compared our temperature profiles reconstructed through Eq. \ref{trec} with the measured temperatures. Given that most of these measurements are limited to the central regions ($<0.5R_{500}$), a similar check cannot be performed for the outer regions. In Appendix \ref{seccomp} we show the temperatures for all 18 objects compared with the available data. Outside the innermost $\sim 1$ arcmin, good agreement is found between the measured and reconstructed temperatures. Our method is typically able to reproduce the observed temperatures with an accuracy of $\leq10\%$. In the innermost regions ($R<0.1R_{500}$), angular resolution effects may play a role, and thus we excluded these regions from our analysis.

\begin{figure}
\resizebox{\hsize}{!}{\includegraphics{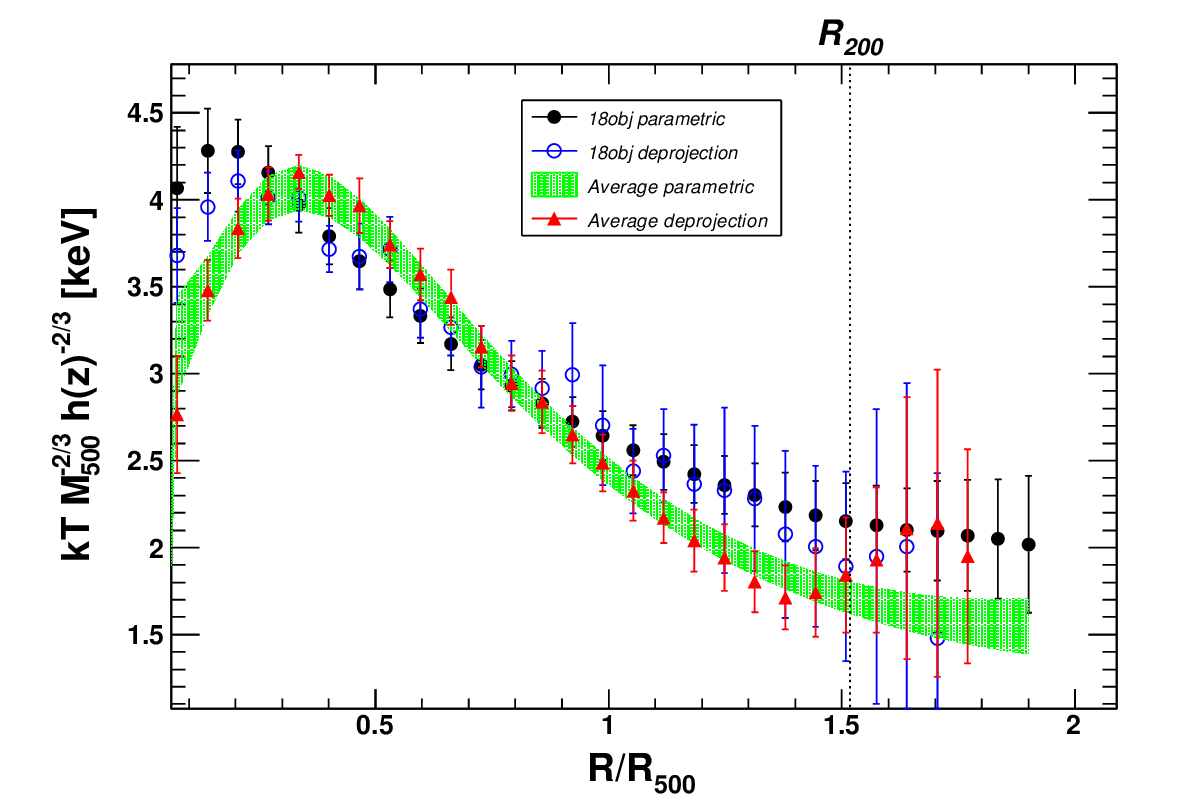}}
\caption{Average self-similar scaled median temperature profile obtained over 18 individual objects compared to the profile derived by combining the universal pressure profile from P12 and the median gas density profile from E12. In both cases, we show the comparison between the profile obtained using the functional form for the gas density (Eq. \ref{vikhform}), with MCMC error bars, and using non-parametric deprojection.}
\label{compt}
\end{figure}

We also tested whether the average profiles extracted from the P12 and E12 samples agree with the individual objects. Indeed, when average quantities are extracted from different samples, selection effects may play a role, in particular in the central regions where large cluster-to-cluster variations are found. In Fig. \ref{compt} we show the temperature profile obtained by computing the median of the temperature profiles from each individual object and compare it with the temperature profiles from the universal pressure profile (P12) and the median gas density profile (E12). The uncertainty in the median was computed at all radii through 1000 bootstrap reshuffling of the populations. At each radius, we randomly reselected a set of 18 data points with repetition, and computed the median of the shuffled dataset. The median value and its error are then given by the median and the root mean square deviation of the distribution obtained in this way. We note that this error estimation method is sensitive to the scatter of the data points around the mean value, i.e., a larger scatter (statistical or intrinsic) leads to a larger error on the median value.

In both cases, we show the temperature profiles obtained from the best-fit parametric function to the EM profile (Eq. \ref{vikhform}) with its associated MCMC error envelope ($1\sigma$ confidence level), and by interpolating the deprojected density profile using a cubic spline method. The difference between these two curves thus highlights the uncertainties associated with the different deprojection techniques. All methods are found to give consistent results around $R_{200}$, where the scatter of the values obtained with the different methods is consistent with zero. In the outer regions, the uncertainties on the median associated with the parametric forms are smaller than with the deprojection. This effect is caused by the larger scatter of the data points in the latter. Indeed, in the parametric approach the shape of the profiles is forced to be regular, which reduces the cluster-to-cluster scatter and thus the errors on the median (see above). For this reason, the profiles obtained through deprojection stop at lower radii.

 In the central regions ($<0.3R_{500}$), the temperature extracted from the average samples falls below the one obtained for the individual systems. This probably indicates the presence of different selection effects in the two samples. Namely, the X-ray sample is likely constituted of objects with a more peaked density profile in the central regions compared to the SZ sample, which leads to a decline in the temperature. In the outer regions, the temperature profiles reconstructed from the average profiles appear to be somewhat steeper than the average of the 18 individual systems. This may once again reflect a selection bias. In all cases, the profiles extracted from the 18 systems should be preferred, since their selection is much better controlled.

For the remainder of the paper, we present only the results obtained with the parametric form for better readability. However, as shown in Fig. \ref{compt}, the profiles obtained through geometrical deprojection are found to agree very well with those obtained with the parametric form, since at all radii the differences are smaller than the statistical uncertainties. This conclusion is also valid for the other derived quantities.

Another possible source of uncertainty in our temperature determination is the conversion between \emph{ROSAT}/PSPC count rate and EM in the low-temperature regime. Indeed, if the temperature in the outer regions falls below $\sim1.5$ keV, this factor depends strongly on the temperature, while at higher temperatures the dependence is mild. To take this effect into account, we used the \emph{Planck} pressure profiles and Eq. \ref{trec} to obtain an estimate of the temperature at each radius assuming temperature profiles from the literature, as in E12. Then we used this temperature to recompute the conversion factors at each radii and extract again the density profiles and the derived quantities. This procedure leads to small differences ($<10\%$) in the conversion factors, which converts into differences of less than 3\% in the reconstructed density. On the other hand, line cooling becomes important especially at low X-ray temperatures, and thus the assumed metallicity can have an impact on the conversion factors. Several works \citep[e.g.,][]{sabrina01,lm08b} indicate that outside the central regions the metallicity profiles are flat with a mean abundance of $\sim0.25Z_\odot$; however, in these works the metallicity profiles do not extend beyond $\sim0.5R_{500}$. Because of the poor spectral resolution of \emph{ROSAT}/PSPC, estimating the behavior of the metallicity profiles in the outer regions is beyond the scope of this paper. However, we note that all the 18 systems in common between \emph{Planck} and \emph{ROSAT} have an average temperature higher than 5 keV, and thus in these systems the temperature is not expected to fall below $\sim2$ keV. Conversely, this effect may affect the average profiles, which include a number of low-temperature systems (see E12).

\subsection{Temperature}
\label{sectemp}

\begin{figure}
\resizebox{\hsize}{!}{\includegraphics{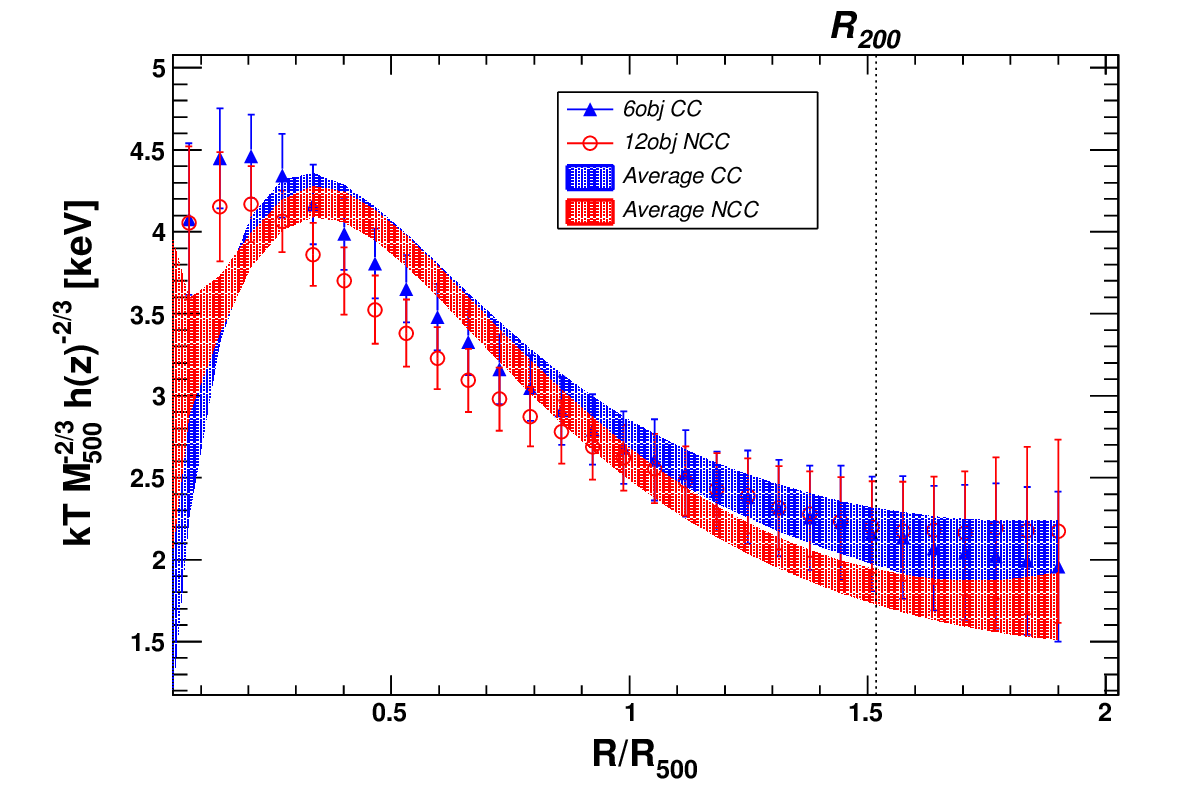}}
\caption{Average self-similar scaled temperature profiles for the CC (blue) and NCC (red) populations separately. The shaded areas show the results obtained by combining the average CC/NCC pressure profiles from P12 with the corresponding profile from E12, while the data points show the median of the individual systems in common between the two samples.}
\label{ktccncc}
\end{figure}

We can see in Fig. \ref{compt} that the combined X-ray/SZ temperature profiles extend out to $\sim2R_{500}$, i.e., out to the virial radius of these systems ($R_{vir}\sim R_{110}$ in a $\Lambda$CDM cosmology), and a factor $\sim2$ beyond the previous results obtained from relatively large cluster samples \citep{lm08,pratt07,vikhlinin06,sabrina,mark98}, thus probing a volume $\sim8$ times larger. The temperature is found to decline by a factor $\sim2.5$ from $0.2R_{500}$ to $R_{200}\sim1.52R_{500}$, and to follow a power-law decrease like $T(r)\sim r^{-\alpha}$, with $\alpha=0.76\pm0.03$ in the radial range $[0.5-2]R_{500}$. A slightly flatter profile is obtained for the median of the 18 individual systems ($\alpha=0.51\pm0.05$). This may indicate that different selection biases are affecting the \emph{Planck} and \emph{ROSAT} samples. In any case, the median data points are found to agree rather well at all radii. Since the pressure and the density are obtained from the same systems, the value obtained by averaging the 18 individual objects is preferred.

\begin{figure}
\resizebox{\hsize}{!}{\includegraphics{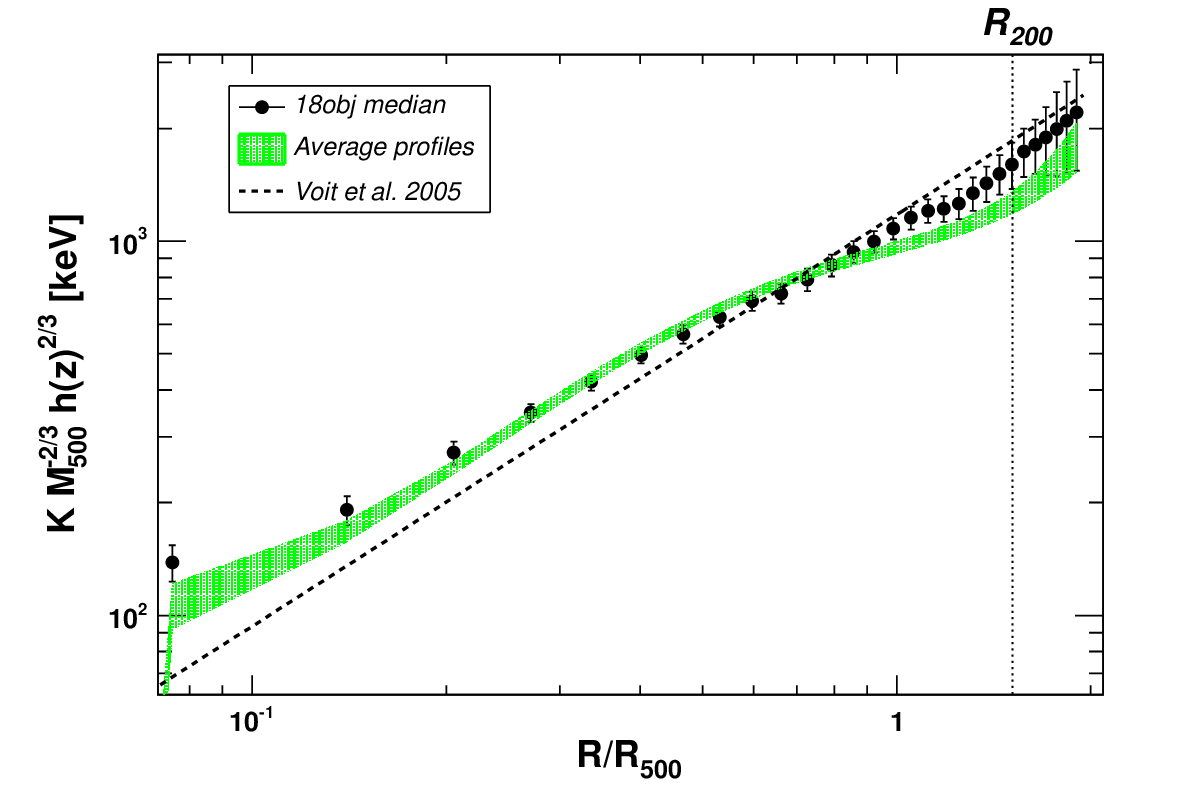}}
\caption{Average self-similar scaled entropy profiles obtained from the median of the 18 individual systems (black points) and the average pressure and density profiles (green shaded area). For comparison, the dashed line shows the relation expected from gravitational collapse only \citep{voit05}, using the proper normalization (Eq. \ref{K500}).}
\label{avK}
\end{figure}

We also computed average temperature profiles for the CC/NCC populations separately, to look for differences between clusters states. The resulting profiles are shown in Fig. \ref{ktccncc}. Both for the average profiles and the individual systems, we observe no significant differences between the two cluster populations. Beyond $\sim0.4R_{500}$, NCC clusters are known to exhibit a higher density (by $\sim$ 15\%) than CC clusters, which was highlighted in E12 \citep[see also][]{maughan11}. When the central slope $\gamma$ of the GNFW profile (Eq. \ref{eqgnfw}) is left free in the MCMC procedure (see Sect. \ref{sec:mcmc}), a similar effect is seen for the pressure, such that the temperature profiles for the two populations are found to agree.

\subsection{Entropy}
\label{secK}

In addition to the gas temperature, we used Eq. \ref{entropy} to probe the behavior of the gas entropy at $R_{500}$ and beyond. Several recent results using the \emph{Suzaku} satellite \citep[e.g.,][]{bautz,kawa,simionescu,walkerpks} found that the gas entropy increases less steeply than expected from gravitational collapse \citep[$K\sim r^{1.1}$,][]{voit05,borgani05}, suggesting that non-gravitational effects \citep[e.g.,][]{lapi} or gas clumping \citep{nagai} play an important role beyond $\sim R_{500}$. Our dataset thus offers an excellent opportunity to test the validity of these results in the framework of a large sample. To allow a meaningful comparison of our entropy profiles with the theoretical expectations, we rescaled them using the self-similar expectation (Eq. \ref{Kssc}) and compared them with the universal profile obtained from non-radiative simulations \citep{voit05}, which is normalized by the quantity

\begin{equation} K_{500}=106 \left(\frac{M_{500}}{10^{14}M_\odot}\right)^{2/3} \left(\frac{1}{f_b}\right)^{2/3} h(z)^{-2/3} \mbox{ keV cm}^2 ,\label{K500}\end{equation}

\noindent where  $f_b=\Omega_b/\Omega_m$ is the cosmic baryon fraction \citep[see Eq. 3 and 4 of][for details]{pratt10}. In Fig. \ref{avK} we show the average entropy profiles obtained from the universal pressure and density profiles, and from the median of the 18 individual systems. The same figure showing the profiles obtained with the two deprojection techniques can be found in Fig. \ref{average_K_all} of the online material. These profiles are compared with the self-similar expectation from \citet{voit05}, scaled according to Eq. \ref{K500} \citep{pratt10,sun09}. As can be seen in Fig. \ref{avK}, the combined X-ray/SZ profiles are indeed flatter than the self-similar relation in the radial range $[0.5-2]R_{500}$. Fitting the average entropy profile with a power law ($K\propto r^{\alpha}$), we find a typical slope $\alpha=0.60\pm0.03$ in this radial range. As already found for the temperature, the individual systems are returning a somewhat steeper slope, $\alpha=0.90\pm0.06$. This may again reflect differences in the selection of the two samples. However, as can be seen in Fig. \ref{avK}, the differences with respect to the \citet{voit05} profile tend to decrease when going farther out. This result disagrees with the behavior generally observed with \emph{Suzaku} \citep[see][and references therein]{badwalker}.

\begin{figure}
\resizebox{\hsize}{!}{\includegraphics{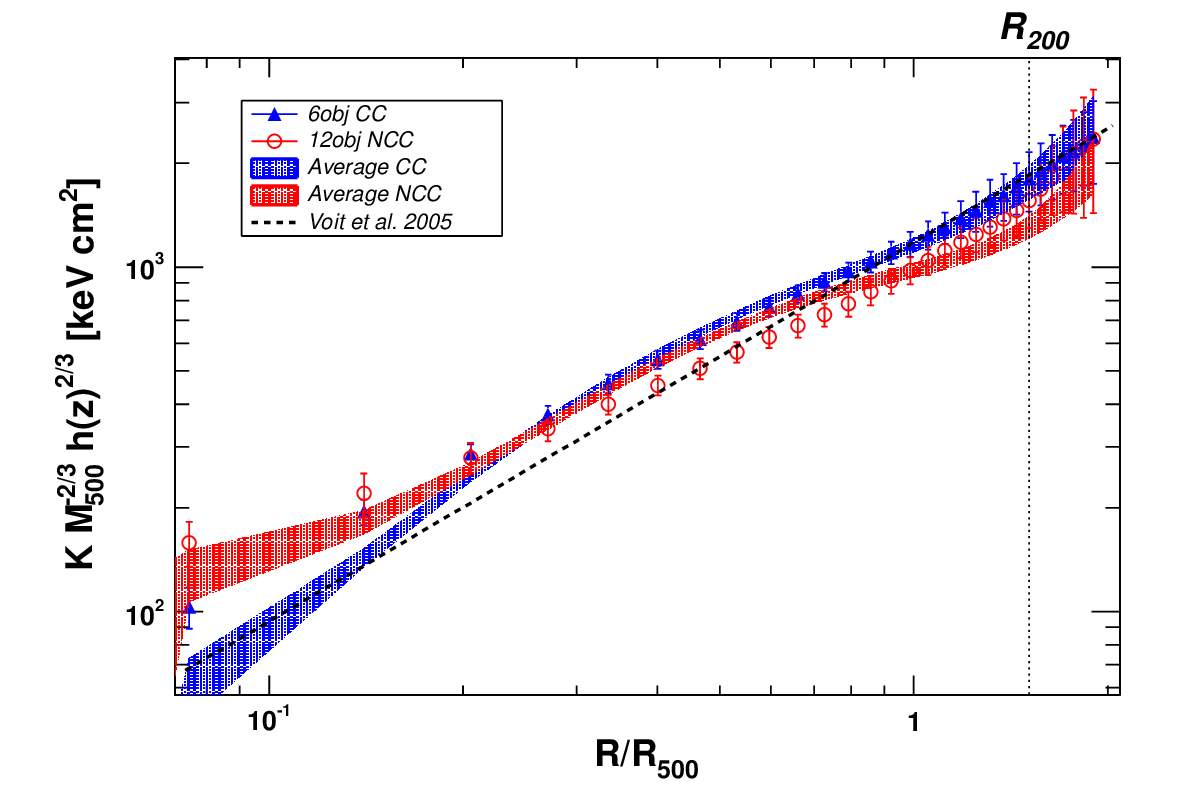}}
\caption{Average entropy profiles for the CC (blue) and the NCC (red) populations. The shaded areas represent the quantities obtained from the average pressure and density profiles, while the data points show the median of the 6 CC and 12 NCC individual systems. For comparison, the dashed line shows the relation expected from gravitational collapse only \citep{voit05}, using the proper normalization (Eq. \ref{K500}).}
\label{kprofs}
\end{figure}

In Fig. \ref{kprofs} we show the average entropy profiles for the CC and NCC populations, derived both from the average pressure and EM profiles and for the 18 individual objects. The profiles are compared to the self-similar expectation $r^{1.1}$. The power-law slopes obtained in the $[0.5-1]$ and $[1-2]R_{500}$ ranges are presented in Table \ref{Kslope}. We notice a significant difference between the CC and NCC slopes in the outer regions. In CC clusters, the slope of the entropy profile is consistent with the self-similar expectations, whereas in NCC clusters we observe slightly flatter entropy profiles.  Similar results are found for the average of the individual systems. This result is explained by the higher density in the outer regions of NCC systems \citep[E12;][]{maughan11}.

\begin{table}
\caption{\label{Kslope}Slope of the average entropy profiles in the radial range $[0.5-1]$ and $[1-2]R_{500}$ obtained from different datasets.}
\begin{center}
\begin{tabular}{lcc}
\hline
Dataset & $\alpha_{[0.5-1]}$ & $\alpha_{[1-2]}$\\
\hline
\hline
Average & $0.59\pm0.06$ & $0.82\pm0.11$\\
18 objects & $0.86\pm0.11$ & $1.01\pm0.21$\\
Average CC & $0.74\pm0.06$ & $1.14\pm0.16$\\
6 objects CC & $0.80\pm0.12$ & $1.16\pm0.24$\\
Average NCC & $0.58\pm0.08$ & $0.85\pm0.11$\\
12 objects NCC & $0.74\pm0.16$ & $1.05\pm0.30$\\
\hline
\end{tabular}
\end{center}
\end{table}

\subsection{Polytropic index}
\label{sec:polytropic}

Another quantity of interest for comparison with semi-analytical models is the polytropic index \citep{tozzi,ascasibar03,capelo}, which is defined as

\begin{equation}\gamma=\frac{{\rm d}\log P}{{\rm d}\log \rho}.\label{eq:polytropic}\end{equation}

\noindent The polytropic relation that relates gas density and pressure by a simple power law is the simplest extension of the isothermal model and is indeed expected in ideal adiabatic processes. In these cases, the polytropic index is equal to 5/3 and represents the ratio of specific heats at constant pressure and constant volume. As a result of the combined action of shock heating and adiabatic compression, the index $\gamma$ is found in numerical studies \citep[e.g.,][]{tozzi} to be between 0.8 and 1.2 over the cluster's volume. It has been shown \citep[e.g.,][]{ascasibar03} that polytropic models provide a fairly accurate description of the radial structure of galaxy groups and clusters, failing in reproducing the cluster properties in merging systems, where the dark matter potential may differ considerably from a predicted NFW form and the assumptions of hydrostatic equilibrium might not hold.

\begin{figure}
\resizebox{\hsize}{!}{\includegraphics{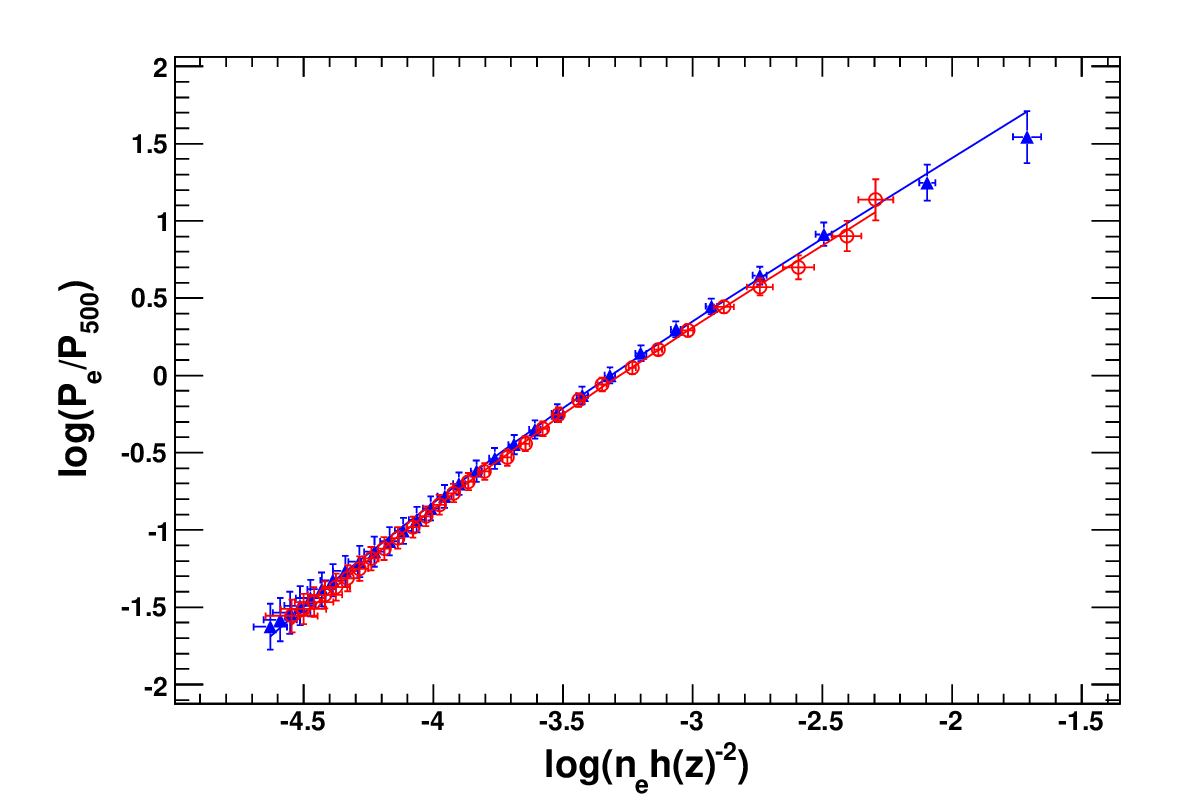}}
\caption{$\log P-\log n$ diagram for the CC (blue triangles) and NCC (red circles) cluster populations. The solid lines show the best fit to the data using a rolling index.}
\label{fig:polytropic}
\end{figure}

Since pressure and density are our primary observables, the dataset explored here is ideal to measure the polytropic index and study its dependence with radius. In Fig. \ref{fig:polytropic} we show the median pressure as a function of the median gas density, for the 6 CC and 12 NCC systems independently. An orthogonal technique was used to fit the data taking the uncertainties in both axes into account. The best-fit values polytropic indices are given in Table \ref{tab:polytropic}.

\begin{table}
\caption{\label{tab:polytropic}Best-fit polytropic index for various datasets and radial ranges.}
\begin{center}
\begin{tabular}{lccc}
\hline
Dataset & $\gamma_{0.1-0.5}$ & $\gamma_{0.5-1}$ & $\gamma_{1-2}$\\
\hline
\hline
Entire sample & $1.06\pm0.07$ & $1.24\pm0.07$ & $1.24\pm0.11$\\
CC & $1.11\pm0.10$ & $1.22\pm0.14$ & $1.23\pm0.16$\\
NCC & $1.05\pm0.12$ & $1.26\pm0.12$ & $1.31\pm0.14$\\
\hline
\end{tabular}
\end{center}
\end{table}

We measure $\gamma = 1.21 \pm 0.03 \;(1.20 \pm 0.03)$ for CC (NCC) systems over the radial range $0.1-2 R_{500}$, with no significative dependence upon the radial range considered. Overall, these values show a significant departure from the isothermal condition in a systematic way at any radius, confirming an effective polytropic index of about $1.2$ predicted as the outcome of the processes of shock heating and adiabatic compression acting on the X-ray emitting plasma during its accretion in the cluster's potential well.

\section{Discussion}

\subsection{Comparison with previous results}

Exploiting the low background of the \emph{Suzaku} satellite, several recent X-ray works claimed the measurement of the thermodynamic properties out to $R_{200}$ \citep[see][and references therein]{akamatsu,badwalker}. These results are reporting a decline in the temperature out to $R_{200}$, in qualitative agreement with the results presented here (see Sect. \ref{sectemp} and Fig. \ref{compt}), and a flattening of the entropy profiles at the outermost radii, falling below the prediction from numerical simulations. For relaxed systems, \citet{badwalker} suggested the existence of a universal entropy profile, showing a general flattening of the entropy profiles. Using a method similar to the one presented here, they combined the universal pressure profile from P12 with the mean deprojected density profile from E12, and claimed that these measurements are supporting the \emph{Suzaku} measurements in the case of relaxed clusters. This claim is however incorrect. Indeed, \citet{badwalker} used the average profiles for the entire P12 and E12 samples, and compared them with results obtained by \emph{Suzaku} on relaxed systems. Our analysis (see Fig. \ref{avK} and \ref{kprofs} and Table \ref{Kslope}) clearly shows that the trend of entropy flattening is driven by the NCC systems, which are in majority both in the \emph{Planck} and \emph{ROSAT} samples. The comparison with the relaxed \emph{Suzaku} systems is therefore not appropriate. Moreover, to compare with the simulation results, \citet{badwalker} arbitrarily set the normalization of the \citet{voit05} profile to match the observed profiles at a radius of $0.3R_{200}$, while the normalization (Eq. \ref{K500}) cannot be fixed arbitrarily \citep{sun09,pratt10,humphrey}. This causes the CC profiles to fall below the self-similar prediction at larger radii, whereas our analysis shows that, in particular for CC clusters, the profiles are in excess compared to the predicted entropy out to $\sim R_{500}$ \citep[see][and references therein]{cavagnolo,pratt10}, but converge to the expected behavior in the outermost regions.

\subsection{Comparison with \texttt{ENZO} simulations}
\label{sec:sim}

\begin{figure*}
\resizebox{\hsize}{!}{\hbox{\includegraphics{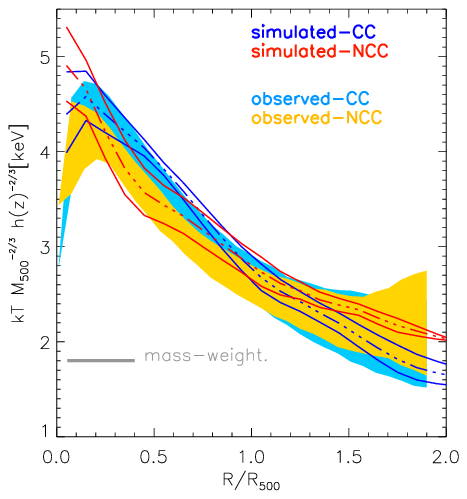}}
\includegraphics{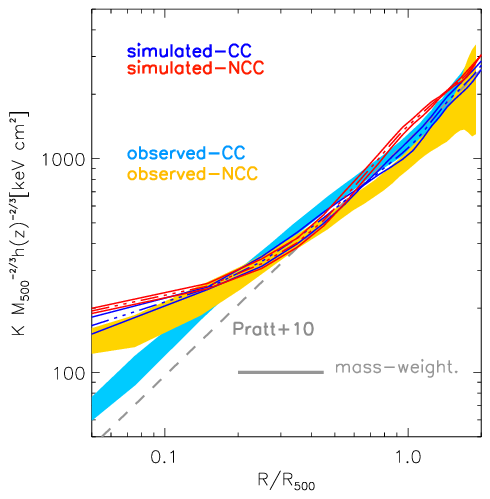}}
\caption{Comparison of average temperature (left) and entropy (right) profiles for the observed CC (cyan) and the NCC (yellow) populations derived from the median of the 18 objects, and the CC-like (blue) and NCC-like (red) populations simulated with \texttt{ENZO}. The dot-dashed lines show the median profiles from our simulated \texttt{ENZO} sample, while the solid lines indicate the confidence intervals on the median. In the right panel, we additionally show the self-similar expectation from \citet{pratt10} (dashed gray line).}
\label{fig:sim}
\end{figure*}

We compared our observational results with the properties of simulated galaxy clusters produced in a set of high-resolution cosmological simulations with the adaptive mesh refinement code \texttt{ENZO} \citep[see][for the description of the simulated cluster sample]{va10kp}. This set of simulations already allowed us to highlight some important differences in the amount of turbulent motions, shock waves, azimuthal scatter, and gas clumping factor between CC-like and NCC-like simulated systems, showing that all these effects tend to be stronger in NCC-like clusters compared to CC-like objects \citep{va10kp,vazza11a,vazzascat,vazza12c,e12}. From the results of E12, we report that the gas density profiles of CC-like and NCC-like clusters with simple non-radiative setup qualitatively agree with observations for $>R_{\rm 500}$, while present results are at odds with observations in the internal radial range. Similar trends are also found in the simulated gas pressure. Overall, these results may be understood in terms of a slightly different gas mass distribution, due to the lack of radiative cooling and inward contraction of the gas/DM atmosphere in non-radiative systems (e.g., E12 and discussion therein). This comparison can now be extended to all other thermodynamic properties of simulated and observed clusters out to $\sim R_{\rm 200}$.

 For each cluster, we computed the profile of mass-weighted temperature, $T_{\rm mw}$, and mass-weighted entropy, $T_{\rm mw}/n^{2/3}$, and generated average profiles for the CC-like and NCC-like classes at $z=0$ using the same weighting by the cluster mass, as in Eq. \ref{tssc}-\ref{Kssc}. Figure \ref{fig:sim} shows the average temperature profiles obtained in this way compared with the observational results of the previous sections (median of the 18 objects). Looking at the mass-weighted temperature for $> 0.1 R_{500}$ (where the effect of missing cooling in the simulations should be unimportant), the degree of qualitative agreement with observations is similar for both populations, with a tendency of simulated CC-like systems to present a slightly steeper trend than the observed CC. Apart from this, the agreement between the observed and simulated temperature profiles is excellent. In the right hand panel of Fig. \ref{fig:sim} we show the average entropy profiles obtained in this way compared with the observational results of the previous sections (median of the 18 objects). While inside $0.2~R_{\rm 200}$ the trend of simulated and observed NCC are qualitatively similar (with NCC systems showing a flatter profile and a higher level of central entropy), the relative difference between CC and NCC is much more marked in real systems, which points toward the additional mechanisms of radiative cooling and non-gravitational heating. For $R>0.2 ~R_{\rm 200}$ the slope of the simulated entropy profiles is close to the self-similar expectation, but with the tendency of producing a $\sim 20-30$ percent higher entropy at all radii. This finding is consistent with the result we obtained in E12, where the gas density at $\sim R_{\rm 200}$ in our simulated clusters has been shown to be slightly lower compared to the density profiles obtained with \emph{ROSAT}. In addition, unlike in observations, the simulated NCC systems present no evidence of a flatter entropy profile at large radii.

\subsection{Entropy flattening or entropy excess?}

In the central regions, galaxy clusters are known to exhibit an excess of entropy with respect to the prediction from pure gravitational collapse, which indicates the existence of an additional heating mechanism, such as feedback from active galactic nuclei (AGN) within already-formed clusters \citep[see][for reviews]{fabian12,mcnamara12} or intense heating episodes until the formation epoch \citep{mccarthy,burns08}. This effect is known to be stronger in group-scale objects \citep[e.g.,][]{david96,ponman99}, which causes the inner slope of the entropy profiles to be flatter in low-mass systems with respect to massive clusters \citep{sun09}. The entropy excess is known to extend significantly beyond the cluster core, at least out to $R_{2500}$ \citep{sun09} and possibly even out to $R_{500}$ \citep{pratt10,humphrey}. The results presented here for relaxed systems indicate that instead of a flattening at large radii, the average entropy profiles actually match the self-similar expectation at $\sim R_{500}$, and are consistent with the expected behavior beyond this radius (see Table \ref{Kslope}). The flatter slopes observed in the outer regions are thus resulting from the entropy excess in the central regions and extending out to large radii, instead of highlighting a further flattening at the outermost radii. 

This result is at odds with most of the results obtained from \emph{Suzaku} data \citep[e.g.,][]{simionescu,akamatsu,walkerpks}. It highlights the importance of using the expected normalization for the self-similar entropy profile when comparing with the predicted behavior instead of normalizing at an arbitrary radius, as shown by \citet{humphrey} for the galaxy group RX J1159+5531, which exhibits a behavior very similar to our findings. Alternatively, the difference could also arise from the fact that most of the \emph{Suzaku} measurements were obtained along narrow sectors, which may not be representative of the global properties in the virial region. Indeed, E12 demonstrated that deviations from azimuthal symmetry increase with radius, in particular in CC systems. A coverage of the entire azimuth is thus required to infer the average properties of a cluster at large radii. In addition, we stress that, unlike the case of \emph{XMM-Newton} \citep{lm08}, the level of systematic uncertainties in recovering the thermodynamical properties at large radii from \emph{Suzaku} data is still poorly known, and no clear criteria are available to assess the reliability of a result (Molendi, private communication). All these effects could contribute to some extent to the discrepancy between the results presented here and those of \emph{Suzaku}. However, determining the exact origin of the discrepancy would require a detailed analysis of each case separately, which is beyond the scope of this paper.

\subsection{Correcting for the gas depletion}

In the REXCESS sample, \citet{pratt10} observed that the excess entropy correlates with the gas depletion with respect to the expected value, such that when the entropy profiles are corrected by the observed gas fraction profile ($(f_{gas}/f_{b})^{2/3}$, where $f_b$ is the expected hot gas fraction), the expectation from gravitational collapse is recovered. In other terms, non-gravitational effects (AGN feedback, preheating) injects a substantial amount of entropy per particle, which leads to a depletion of gas in the central regions and a high central entropy. Correcting for the gas depletion allows one to correct for the redistribution of the gas induced by non-gravitational effects. 

\begin{figure}
\resizebox{\hsize}{!}{\includegraphics{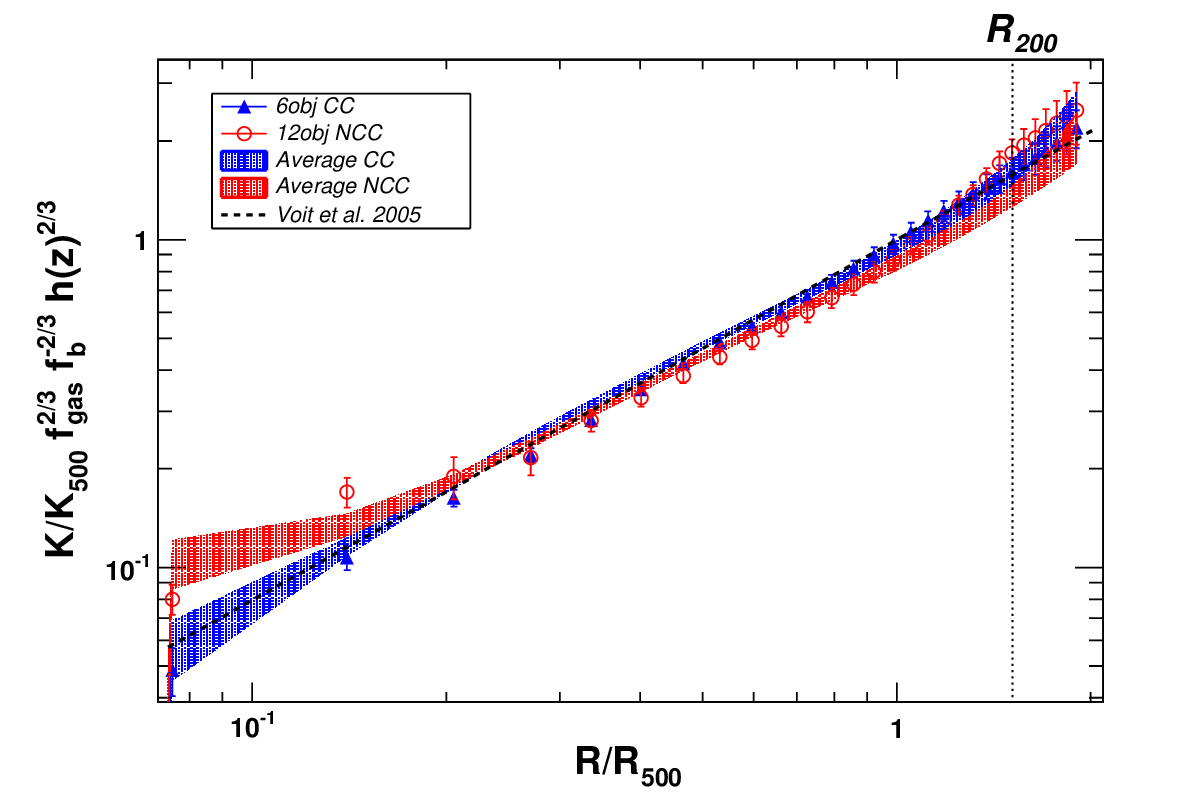}}
\caption{Average self-similar scaled entropy profiles for the CC (blue) and NCC (red) cluster populations, corrected for the variation of $f_{gas}$ with respect to the expected value \citep{pratt10}. The dashed line shows the entropy profile expected from gravitational collapse only \citep{voit05}.}
\label{K_fgcorr}
\end{figure}

The dataset explored here offers the unique opportunity to test whether this effect also applies to the external regions. Using the gas fraction profiles presented in Paper II, we rescaled the reconstructed entropy profile using the method introduced by \citet{pratt10}, both for the average profile and the 18 individual systems. In Fig. \ref{K_fgcorr} we show the entropy profiles scaled by the self-similar $K_{500}$ (Eq. \ref{K500}) and corrected for the variations in gas fraction following \citet{pratt10}. As we can see in the figure, the result obtained within $R_{500}$ by \citet{pratt10} holds over the entire radial range \citep[see also][]{humphrey}, in particular in the case of relaxed CC clusters. For NCC clusters, slight deviations with respect to the prediction are observed, although the general trend is qualitatively similar. In these systems, we observe a gas fraction in excess of the expected hot gas value beyond $R_{500}$ (see Paper II), which can be caused either by the breakdown of hydrostatic equilibrium under the influence of non-gravitational effects (e.g., turbulence, bulk motions, cosmic rays), or by an inhomogeneous gas distribution (see Paper II and references therein). Correcting for this observed excess of baryons in cluster outskirts thus allows one to recover the expected behavior in the outer regions of perturbed systems. Conversely, in CC systems both the observed hot gas fraction and the entropy (Fig \ref{kprofs}) agree with the expected behavior, indicating that on the large scale these systems are governed mainly by simple gravitational physics. 

\subsection{Possible interpretations}

\subsubsection{Gas clumping}

As we can see in Fig. \ref{kprofs}, the entropy of NCC clusters falls below the expectation from pure gravitational collapse beyond $\sim 0.7R_{500}$. Since the X-ray emissivity is proportional to the squared density, inhomogeneities in the gas distribution can lead to biased density measurements if one assumes a constant density within each spherical shell. This could bias the measured entropy low, which would lead to a deficit of entropy with respect to the expected behavior \citep{mathiesen,nagai,vazza12c}. The lower entropy of NCC systems could thus indicate the presence of significant gas clumping at large radii. In the same way, large deviations from azimuthal symmetry could affect the NCC population, although the mean azimuthal scatter on large scales appears to be similar in both cluster populations (see E12).

Following \citet{mathiesen}, we define the clumping factor as the bias in the density reconstructed from X-ray measurements compared to the true density, $C=\langle\rho^2\rangle/\langle\rho\rangle^2$. Assuming that the difference between the NCC entropy profile and the \citet{voit05} profile is caused by gas clumping, one can thus derive the clumping factor needed to recover the self-similar expectation. At $R_{200}$, we find $\sqrt{C}=1.23\pm0.06$ ($1.18\pm0.12$ from the 12 individual systems). This value is in the range allowed by simulations \citep{mathiesen,nagai}, and well below the value proposed by \citet{simionescu} to explain the \emph{Suzaku} observations of the NW arm of Perseus. This suggests that either this measurement, which covers less than 5\% of the azimuth at $R_{200}$, is not representative of the cluster as a whole, or another mechanism must be invoked to reconcile it with the theoretical expectations.

Numerical simulations also predict differences in the clumping factor between merging systems and relaxed clusters, even if by a smaller factor than required here. Since merging clusters exhibit more accretion than relaxed systems, especially along filaments, they are expected to host a significantly larger number of clumps \citep{nagai,vazza12c}, in agreement with the trend observed here for NCC. Conversely, in the case of CC systems good agreement is found with the predictions of gravitational collapse of uniform smooth cells of matter, which translates into a clumping factor consistent with unity at all radii.

\subsubsection{Non-thermal energy supply}

Cosmological simulations generally agree in reporting that the amount of turbulent energy is expected to be $\sim 20-30$ percent of the thermal energy at $R_{\rm 200}$ in NCC systems, and only $\sim 15$ percent in CC \citep[e.g.,][]{lau09,vazza11a}. After major merging events, the magnetic energy in the ICM should be boosted by a factor $\sim 2-4$ \citep[e.g.,][]{roettiger,dolag02,xu}, and so is the relative boost of CR energy within 
clusters following  merger shock waves \citep[e.g.,][]{pfrommer07,vazza12a}. Indeed, the more intense activity of merger and large-scale accretions (e.g. from filaments) that is likely at work in NCC systems is expected to promote a larger injection of turbulent motions, cosmic-ray (CR) energy and magnetic field amplification compared to CC systems, soon after crossing the cluster virial radius. All these mechanisms reduce the thermalization efficiency of kinetic energy within clusters so that, for a given total mass, an NCC system is expected to have a lower temperature than a CC system in the outskirts $(<R_{200})$,  where non-thermal phenomena are expected to be more important due to  the departure from the simple ``smooth accretion" at work in real systems. This is indeed what various simulations report \citep{lau09,vazza11a}, including those presented in this paper (see Sect. \ref{sec:sim}).

Conversely, our observations show that, outside the core, the CC and NCC temperature profiles are indistinguishable  (see Fig. \ref{ktccncc}). This would seem to imply that the thermalization process in the outer regions of NCC and CC systems is equally efficient, possibly indicating that the mechanism that are responsible for thermalization are significantly more rapid than those currently included in simulations. While this is an intriguing possibility, we must however note that, given the limited statistical quality of our data, we cannot exclude that NCC profiles are on average lower than CC profiles by about 20-25\% at $R_{200}$. Thus, while our data do not require that NCC and CC system have a different fraction of non-thermal energy they cannot, at least to some extent, exclude this. 

\section{Conclusion}

Combining X-ray (\emph{ROSAT}) and SZ (\emph{Planck}) data, we analyzed the properties of the ICM at large radii using both sample-averaged profiles and the data for 18 individual systems. This allowed us to measure for the first time the thermodynamic properties (temperature, entropy) of galaxy clusters out to the virial radius in relatively large samples. Gravitating mass profiles obtained using the same dataset are presented in a companion paper. Our results can be summarized as follows:

\begin{itemize}
\item
We have shown that the thermodynamic properties of the ICM can be efficiently recovered by using the thermal pressure profiles from SZ observations and the gas density profiles from X-ray data. Comparing the results with temperature profiles from the literature, we found that this method allows us to reconstruct the gas temperature within $\sim10\%$ of the X-ray spectroscopic value.

\item
In cluster outskirts, the temperature is found to decrease like $T(r)\propto r^{-0.51\pm0.06}$ in the radial range $[0.5-2]R_{500}$ ($T(r)\propto r^{-0.76\pm0.03}$ from the average profiles). The temperature at $R_{200}$ is on average lower than that at $0.2R_{500}$ (which is close to the spectroscopic X-ray temperature) by a factor of 2.5. No significant differences are observed between the temperature profiles of relaxed CC clusters and unrelaxed NCC objects.

\item
As indicated by several recent \emph{Suzaku} observations, the entropy $K=Pn_e^{-5/3}$ is slightly flatter than the self-similar expectation in the radial range $[0.5-2]R_{500}$, although it rises steadily with radius. However, we note that at the outermost radii the observed behavior of the entropy is close to the one expected from simulations, once the proper normalization for the entropy profile is used \citep{voit05,pratt10}. When a correction for the gas depletion is introduced \citep{pratt10}, the entropy profiles for the two populations are found to be in close agreement with the expectation from gravitational collapse. Therefore, instead of a decrement of entropy at large radii as indicated by \emph{Suzaku} \citep[see also][]{badwalker}, our results indicate that the well-known entropy excess observed in the inner regions extends well beyond the core of clusters out to $R_{200}$. 

\item
Unlike what was recently reported by \citet{badwalker}, the average entropy profile in relaxed systems is found to agree with the expectations from gravitational collapse at the outermost radii, while in perturbed NCC systems the average entropy profile falls below the self-similar prediction. This indicates that non-gravitational effects (turbulence, cosmic rays) and/or accretion of non-virialized structures (i.e., gas clumping) is more important in these systems. 

\item
The polytropic index is found to exhibit a weak radial dependence and to have an average value of 1.2, with no obvious difference between the two cluster populations. This result agrees well with the theoretical expectations \citep[e.g.,][]{tozzi,capelo}.

\item
The differences observed between CC and NCC clusters at large radii could be explained by a higher clumping factor in the latter population. If the deviations with respect to the self-similar expectations are due to gas clumping, we measure a clumping factor $\sqrt{C}=1.23\pm0.06$ at $R_{200}$ for NCC systems, which is well within the range expected from numerical simulations. In CC systems, our observations are consistent with a smooth ICM at all radii.

\item
Alternatively, some of the observed trends in CC and NCC profiles could be interpreted in terms of a more efficient injection of non-thermal energy (turbulence, cosmic rays, magnetic fields) in the latter population following from merging events, at the level of $\sim 20-40$ percent at $R_{\rm 200}$. However, given the lack of observed differences in temperature between CC and NCC systems, this interpretation is not favored by our data, although it cannot be firmly excluded.

\end{itemize}

\acknowledgements{We thank Etienne Pointecouteau for useful comments, which helped us to improve the paper. F.V. acknowledges the usage of computational resources under the CINECA-INAF 2008-2010 agreement. F.V. gratefully acknowledges C. Gheller and G. Brunetti for their contribution to produce the cluster catalog used in this work.}

\bibliographystyle{aa}
\bibliography{thermo}

\normalsize
\appendix

\section{Master table}

In Table \ref{tab1} we give the list of clusters in common between the \emph{Planck} and \emph{ROSAT} samples. For the detailed properties of these clusters, we refer to \citet{planck11} and E12.

\begin{table*}
\begin{center}
\caption{\label{tab1}Individual systems in common between the E12 and P12 samples.}
\begin{tabular}{ccccccc}
\hline
Name & State & $\alpha_T$ & $\alpha_K$ & $f_{gas,500}$ & $f_{gas,200}$ & $R_{\max}$ \\
\hline
\hline
A85 & CC & $0.05\pm0.08$ & $1.52\pm0.09$ & $11.3\pm1.1$ & $9.4\pm1.0$ & 1.59\\
A119 & NCC & $-0.41\pm0.13$ & $0.94\pm0.14$ & $15.3\pm2.4$ & $13.9\pm2.5$ & 1.55\\
A401 & NCC & $-0.27\pm0.08$ & $1.02\pm0.08$ &  $15.8\pm1.7$ & $18.9\pm2.1$ & 1.83\\
A478 & CC & $-0.40\pm0.13$ & $0.94\pm0.13$ & $13.4\pm1.7$ & $15.7\pm2.3$ & 1.81\\
A665 & NCC & $-0.54\pm0.10$ & $0.86\pm0.11$ & $20.7\pm2.6$ & $24.8\pm3.4$ & 2.11\\
A1651 & NCC & $-0.09\pm0.10$ & $1.35\pm0.10$ & $12.8\pm1.3$ & $11.8\pm1.5$ & 1.43\\
A1689 & NCC & $-0.66\pm0.11$ & $0.72\pm0.12$ & $13.9\pm1.9$ & $20.9\pm3.3$ & 2.10\\
A1795 & CC & $-0.60\pm0.14$ & $0.95\pm0.15$ & $10.9\pm1.4$ & $12.3\pm2.2$ & 1.56\\
A2029 & CC & $-0.44\pm0.09$ & $0.97\pm0.10$ & $13.8\pm1.5$ & $15.8\pm1.8$ & 1.62\\
A2163 & NCC & $-0.29\pm0.09$ & $1.15\pm0.11$ & $32.5\pm3.0$ & $27.3\pm4.0$ & 1.53\\
A2204 & CC & $-0.61\pm0.09$ & $0.70\pm0.11$ & $14.9\pm1.7$ & $20.1\pm3.2$ & 1.80\\
A2218 & NCC & $-0.16\pm0.10$ & $1.10\pm0.10$ & $19.6\pm1.6$ & $22.6\pm2.5$ & 2.07\\
A2255 & NCC & $-0.54\pm0.10$ & $0.72\pm0.10$ & $11.7\pm1.4$ & $15.5\pm2.3$ & 2.11\\
A2256 & NCC & $-1.02\pm0.13$ & $0.42\pm0.10$ & $14.2\pm3.0$ & $21.7\pm4.4$ & 1.81\\
A3112 & CC & $-0.80\pm0.10$ & $0.45\pm0.10$ & $11.8\pm1.3$ & $16.5\pm2.3$ & 1.75\\
A3158 & NCC & $-0.57\pm0.12$ & $0.77\pm0.13$ & $17.5\pm2.6$ & $22.8\pm4.0$ & 1.68\\
A3266 & NCC & $-0.22\pm0.12$ & $1.64\pm0.14$ & $12.3\pm1.7$ & $10.0\pm1.9$ & 1.51\\
A3558 & NCC & $-0.14\pm0.10$ & $1.34\pm0.11$ & $13.9\pm1.7$ & $12.3\pm1.6$ & 1.66\\
\hline
\end{tabular}
\end{center}
Column description. 1. Core state: cool-core (CC) or non-cool core (NCC).  2. Slope of the temperature profile in the radial range $[0.5-1.5]R_{500}$. 3. Same for the entropy. 4. Gas fraction at $R_{500}$ (see Paper II). 5. Same at $R_{200}$. 6. $2\sigma$ maximum detection radius in the \emph{ROSAT} data.
\end{table*}

\section{Comparison with literature temperature profiles}
\label{seccomp}

To ensure that the reconstruction of thermodynamic quantities from the X-ray density and the SZ pressure is giving correct results, we used the data for the 18 individual systems in common between the \emph{ROSAT} and \emph{Planck} samples (see Sect. \ref{formalism} and Table \ref{tab1}) and compared them with temperature profiles from \emph{XMM-Newton} \citep{xmmcat} or \emph{Chandra} \citep{cavagnolo}. Given that the angular resolution of \emph{ROSAT} is poorer than that of \emph{Chandra} and \emph{XMM-Newton}, we restricted our comparison to the radial range beyond $0.1R_{500}$. We also excluded the observational data points beyond $R_{500}$, since the techniques used to extract X-ray temperatures in large samples may not be valid in regions with a low signal-to-background ratio \citep[see][]{ettoriwfxt}.

\begin{figure}
\resizebox{\hsize}{!}{\includegraphics{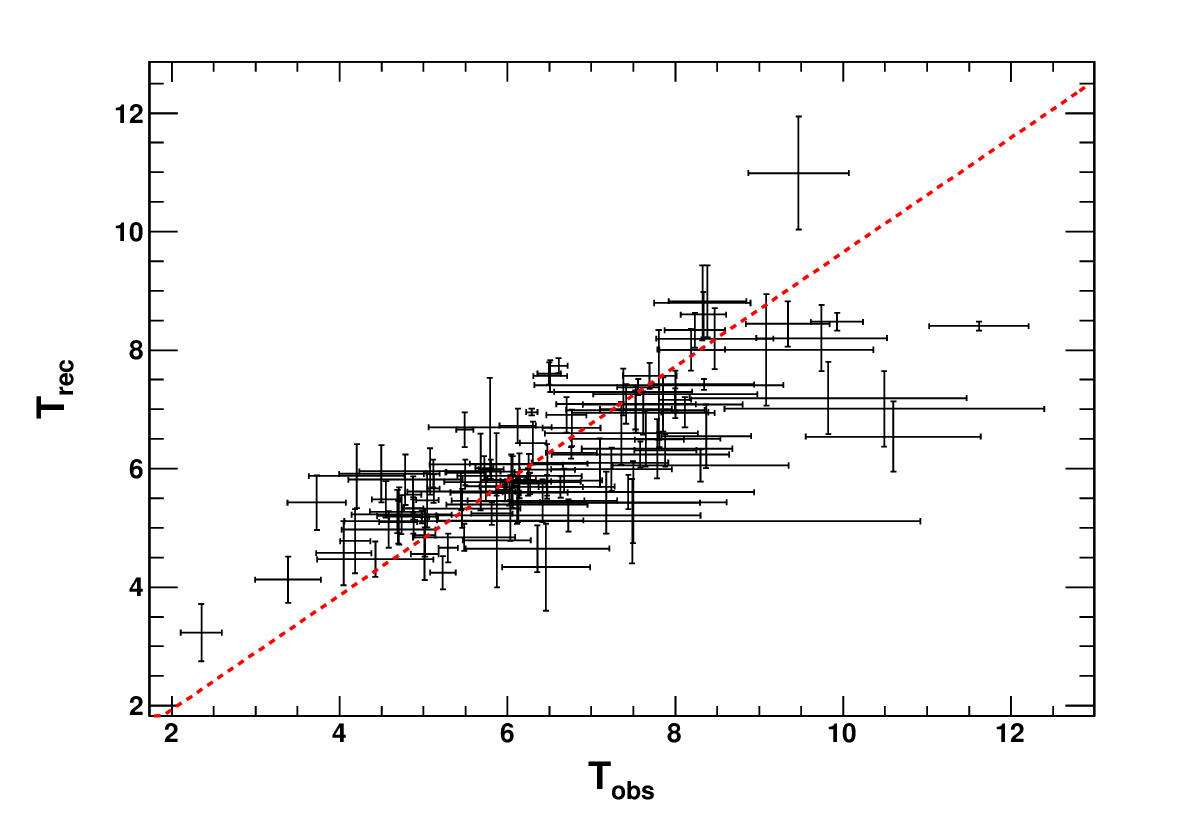}}
\caption{\emph{Planck}/\emph{ROSAT} reconstructed temperature as a function of the observed temperature from \emph{XMM-Newton} \citep{xmmcat} or \emph{Chandra} \citep{cavagnolo} for the 18 objects in common between the two samples. The dashed red curve represents the best-fit function $T_{rec}=0.97T_{obs}$.}
\label{figtrec}
\end{figure}

In Fig. \ref{figtrec} we show the temperatures reconstructed using our method versus the observed temperatures for all available data points. The data were fit with a maximum-likelihood estimator that accounts for the scatter around the mean value, which needs to be computed simultaneously \citep[see][and paper II]{maccacaro}. The best-fit value for the data points in the $[0.1-1]R_{500}$ range gives

\begin{equation}\frac{T_{rec}}{T_{obs}}=0.97\pm0.01,\end{equation}

\noindent with a scatter of 9\%. This analysis demonstrates that combining X-ray density profiles with SZ pressure profiles is an effective way of reconstructing thermodynamic quantities.

\section{Notes on individual objects}

\begin{itemize}
\item
\textit{\large{A478:}} The line of sight of this system is known to exhibit a complex, varying column density \citep{pointe478}. For this work, as in E12, we fixed $N_H$ to the 21cm value \citep{kalberla}. This may result in a slightly incorrect density profile.\\

\item
\textit{\large{A401:}} This system is in an early stage of merging with its neighbor A399, located $\sim35^\prime$ SW, and connected to it by a bridge of X-ray emission. The density profile was thus extracted in a sector excluding the direction connecting A401 to A399. Because it is unclear whether the same was done for the SZ profile, the resulting profiles may be unreliable.\\

\item
\textit{\large{A2163:}} This very perturbed merging cluster may be out of hydrostatic equilibrium, which causes the mass estimates from X-rays to be overestimated by a factor of $\sim2$ with respect to the mass derived from weak lensing and galaxy distribution \citep{soucail}. Therefore, the scaling factors (Eq. \ref{p500} and \ref{fofm}) may be incorrect, as well as the estimate of $R_{500}$ from \citet{planck11}.\\

\end{itemize}

\Online

\begin{figure*}
\resizebox{\hsize}{!}{\vbox{\includegraphics{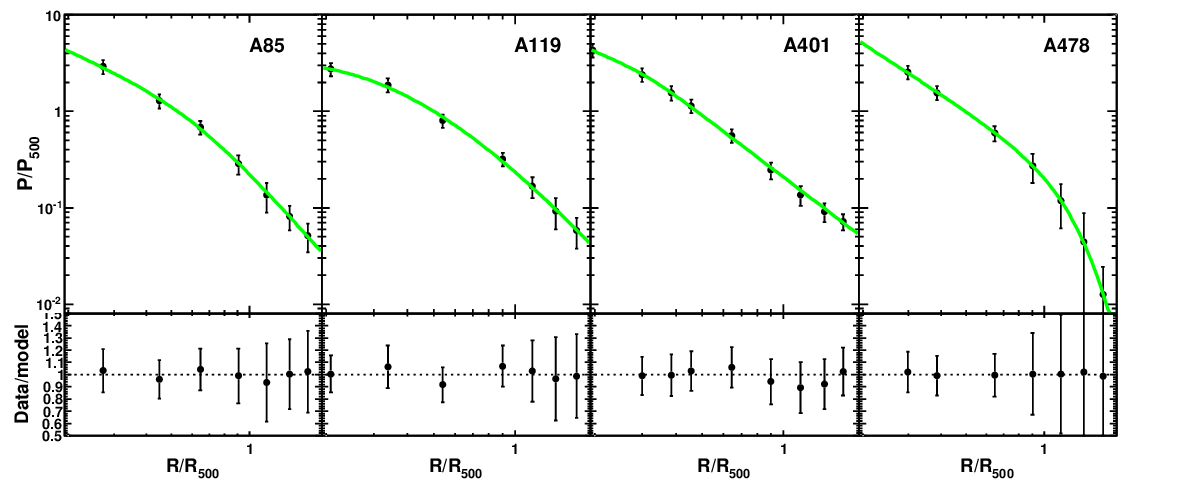}
\includegraphics{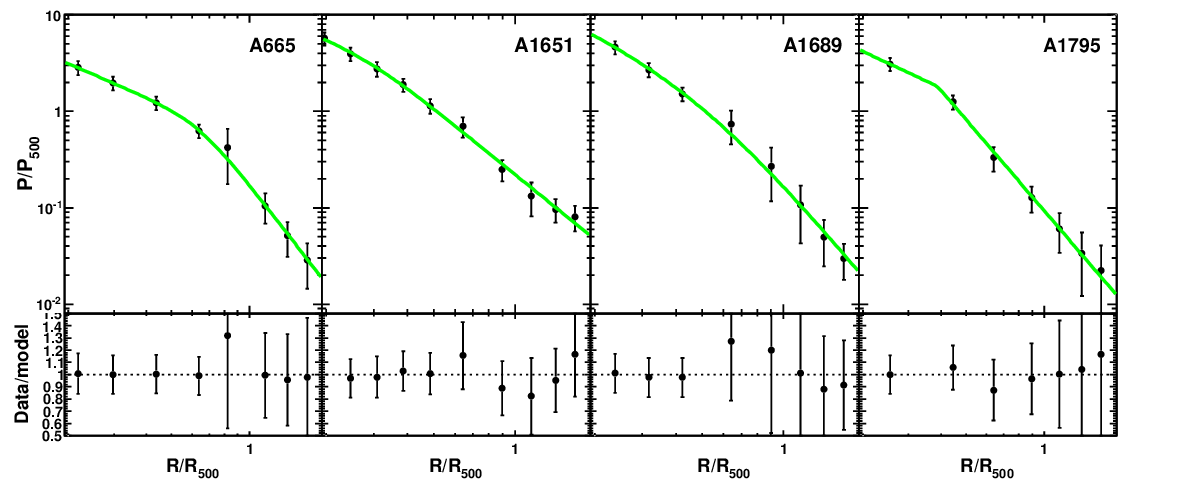}
\includegraphics{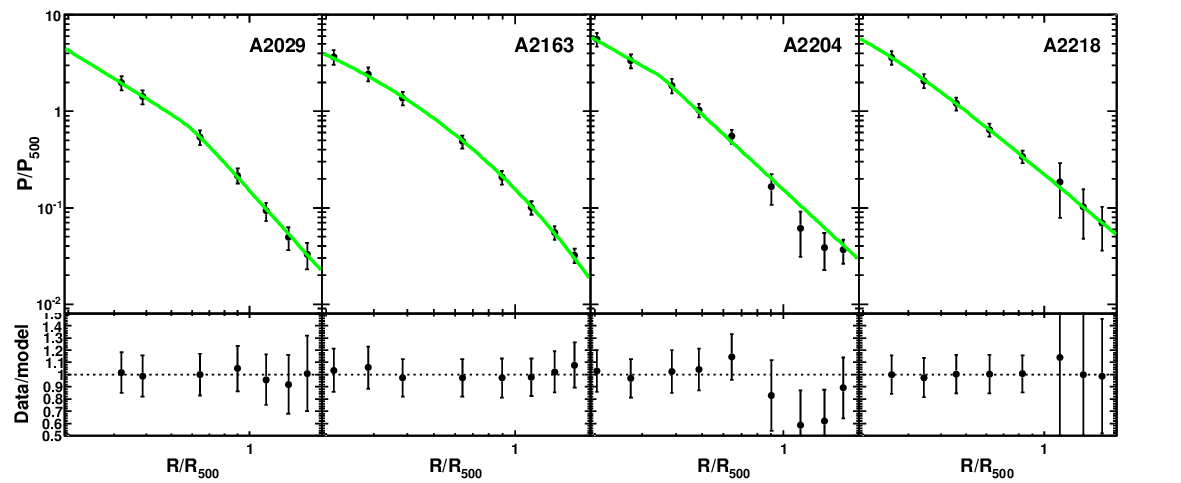}
}}
\caption{Best fit to the \emph{Planck} pressure profiles of the 18 systems (black points, from P12) with a GNFW profile (green lines), in the radial range of interest for this paper ($[0.2-1.9]R_{500}$). The bottom panels show the ratio between the data and the model.}
\label{planckprofs}
\end{figure*}

\begin{figure*}
\resizebox{\hsize}{!}{\vbox{\includegraphics{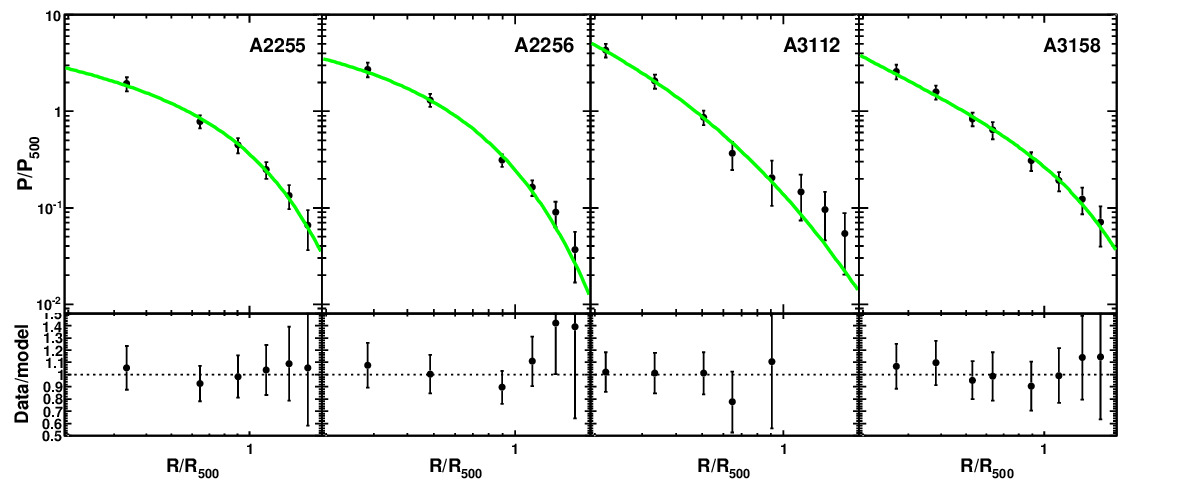}
\includegraphics{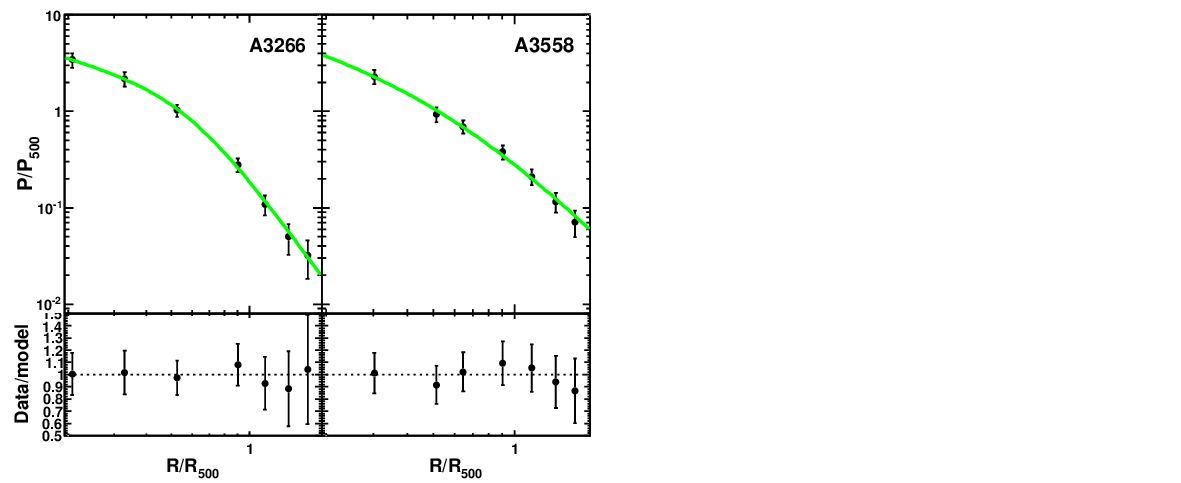}
}}
\caption{Same as Fig. \ref{planckprofs}.}
\label{planckprofs2}
\end{figure*}

\begin{figure*}
\resizebox{\hsize}{!}{\vbox{\includegraphics{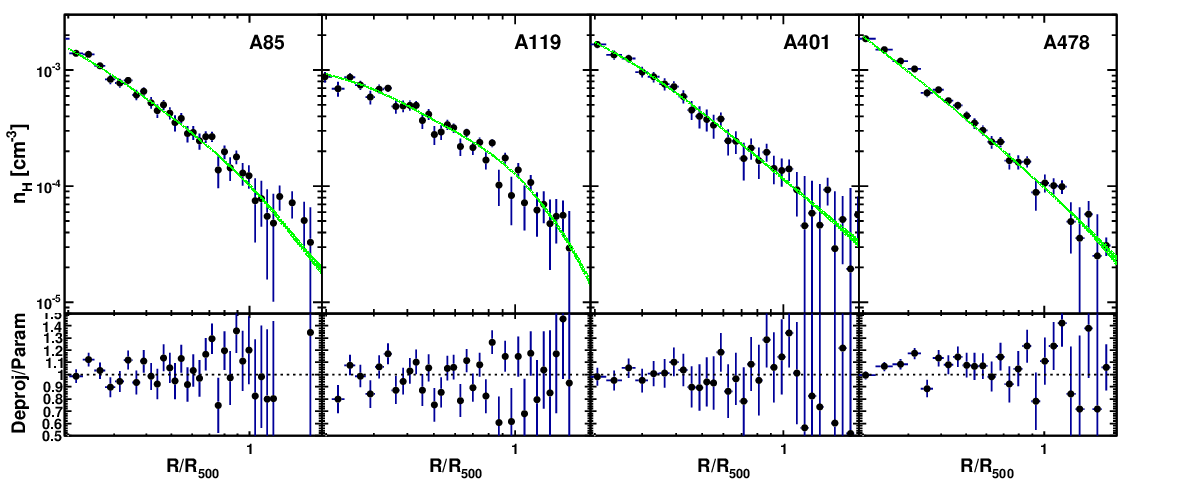}
\includegraphics{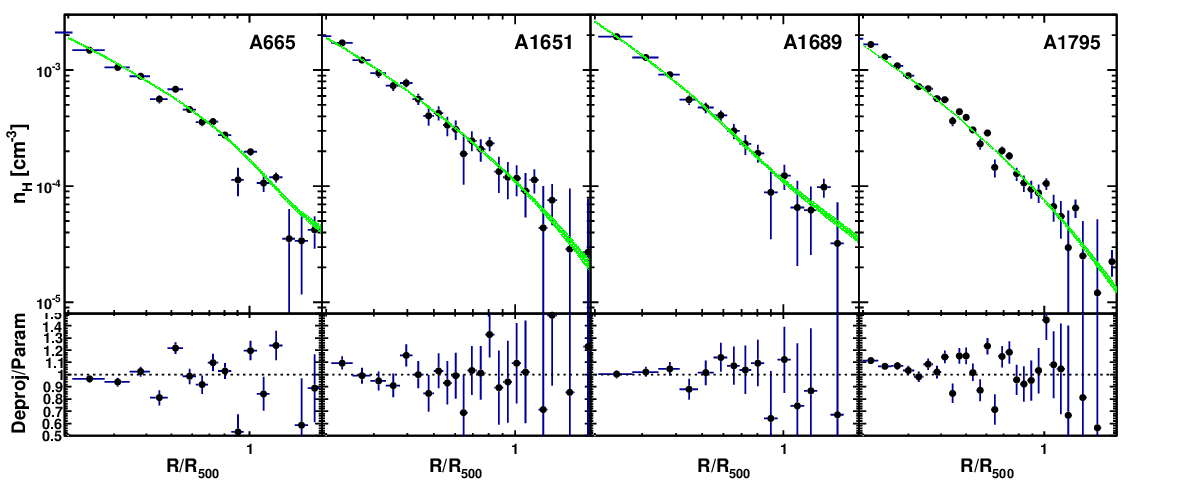}
\includegraphics{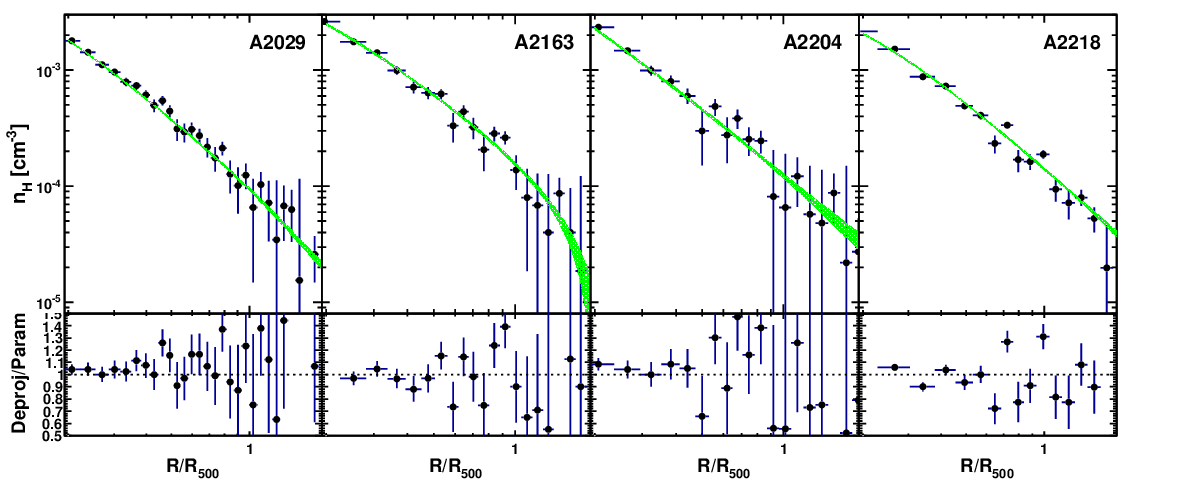}
}}
\caption{\emph{ROSAT} density profiles of the 18 systems obtained through geometrical deprojection (black points) and parametric forward fitting (green-shaded areas), in the radial range of interest for this paper ($[0.2-1.9]R_{500}$). The bottom panels show the ratio between the two methods.}
\label{rosatprofs}
\end{figure*}

\begin{figure*}
\resizebox{\hsize}{!}{\vbox{\includegraphics{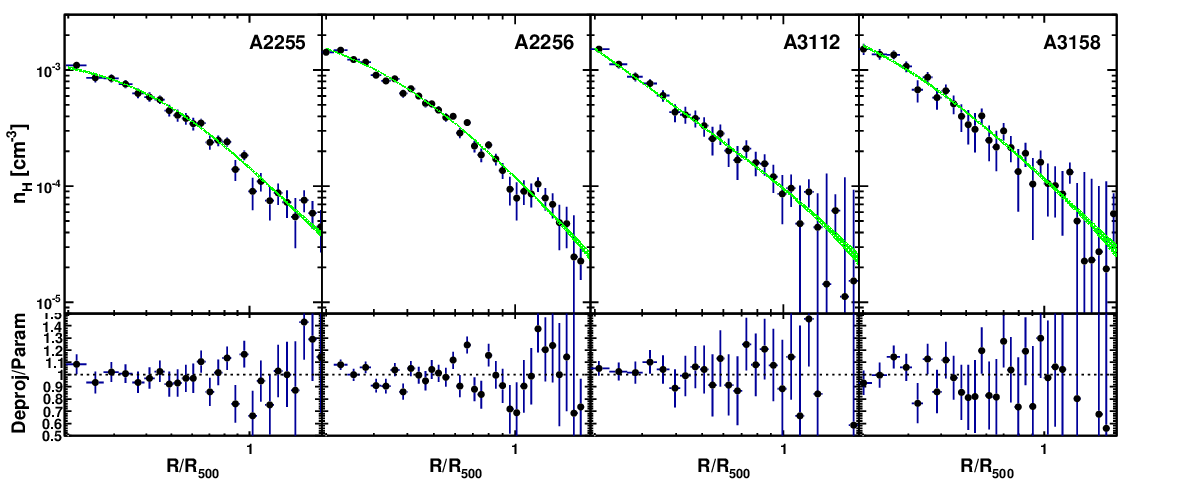}
\includegraphics{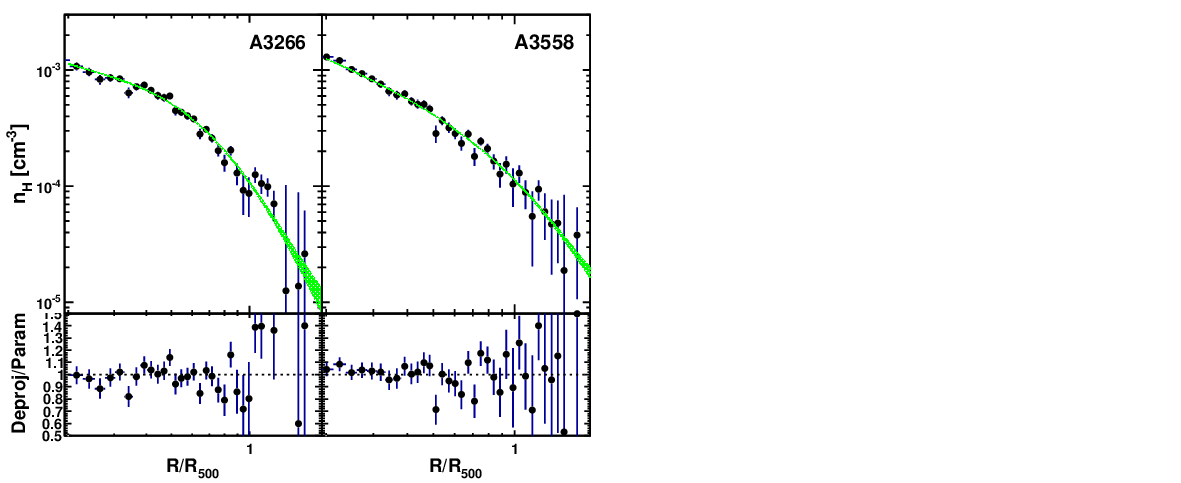}
}}
\caption{Same as Fig. \ref{rosatprofs}.}
\label{rosatprofs2}
\end{figure*}

\begin{figure*}
\resizebox{\hsize}{!}{\vbox{\includegraphics{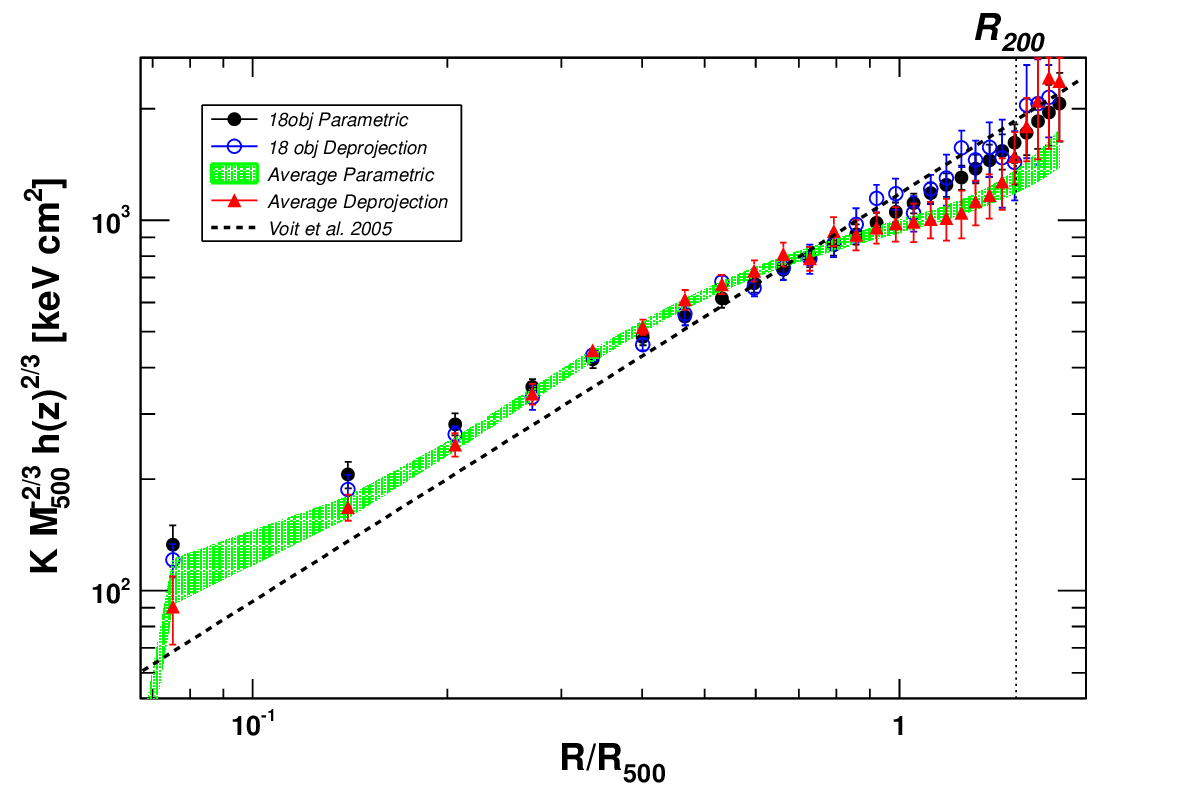}}}
\caption{Average entropy profiles for the total population for the average populations and for the median of the 18 systems in common. Similar to Fig. \ref{avK}, but showing the differences between the parametric and deprojection techniques in both cases.}
\label{average_K_all}
\end{figure*}


\end{document}